\documentclass[journal]{IEEEtran}

\usepackage{cite}
\usepackage{amsmath,amssymb,amsfonts}
\usepackage{tabularx}
\usepackage{booktabs} % For better table rules
\usepackage{subcaption} % 用于子图和子表
\usepackage{graphicx}
\usepackage[font=footnotesize, labelfont=bf]{caption}
\usepackage{textcomp}
\usepackage{xcolor}
\usepackage{algorithm}
\usepackage{algpseudocode}
\usepackage{enumitem}
\begin{document}

\title{Demand-Aware Beam Hopping and Power Allocation 
 for Load Balancing in Digital Twin empowered LEO Satellite Networks}

\author{Ruili~Zhao,~\IEEEmembership{Student Member,~IEEE,}
        Jun~Cai,~\IEEEmembership{Senior Member,~IEEE,}
        Jiangtao~Luo,~\IEEEmembership{Senior Member,~IEEE,} \\ Junpeng~Gao,~\IEEEmembership{Student Member,~IEEE} and Yongyi~Ran,~\IEEEmembership{ Member,~IEEE,}
% <-this % stops a space
\thanks{This work is jointly supported by National Science Foundation of China (No.62171072,62172064,62003067), Chongqing Postgraduate Research and Innovation Project (CYB21204) and the China Scholarship Council program (Project ID: 202308500242). (Corresponding authors: Jun Cai).}% <-this % stops a space
\thanks{Ruili Zhao and Junpeng Gao are with the School of Communication and Information Engineering, Chongqing University of Posts and Telecommunications, Chongqing 400065, China (e-mail: d200101029@stu.cqupt.edu.cn; d200101007@stu.cqupt.edu.cn).}% <-this % stops a space
\thanks{Jiangtao Luo and Yongyi Ran are with the Electronic Information and Networking Research Institute, Chongqing University of Posts and Telecommunications, Chongqing 400065, China (e-mail: luojt@cqupt.edu.cn; ranyy@cqupt.edu.cn).}% <-this % stops a space
\thanks{Jun Cai is with the Network Intelligence and Innovation Laboratory (NI2L), Department of Electrical and Computer Engineering, Concordia University, Montreal, QC H3G 1M8, Canada (e-mail: jun.cai@concordia.ca).}% <-this % stops a space
}

% The paper headers
%\markboth{Journal of \LaTeX\ Class Files,~Vol.~14, %No.~8, August~2015}%
%{Shell \MakeLowercase{\textit{et al.}}: Bare Demo of %IEEEtran.cls for IEEE Journals}
\maketitle
\begin{abstract}
Low-Earth orbit (LEO) satellites utilizing beam hopping (BH) technology offer extensive coverage, low latency, high bandwidth, and significant flexibility. However, the uneven geographical distribution and temporal variability of ground traffic demands, combined with the high mobility of LEO satellites, present significant challenges for efficient beam resource utilization. Traditional BH methods based on GEO satellites fail to address issues such as satellite interference, overlapping coverage, and mobility. This paper explores a Digital Twin (DT)-based collaborative resource allocation network for multiple LEO satellites with overlapping coverage areas. A two-tier optimization problem, focusing on load balancing and cell service fairness, is proposed to maximize throughput and minimize inter-cell service delay. The DT layer optimizes the allocation of overlapping coverage cells by designing BH patterns for each satellite, while the LEO layer optimizes power allocation for each selected service cell. At the DT layer, an Actor-Critic network is deployed on each agent, with a global critic network in the cloud center. The A3C algorithm is employed to optimize the DT layer. Concurrently, the LEO layer optimization is performed using a Multi-Agent Reinforcement Learning algorithm, where each beam functions as an independent agent. The simulation results show that this method reduces satellite load disparity by about 72.5\% and decreases the average delay to 12ms. Additionally, our approach outperforms other benchmarks in terms of throughput, ensuring a better alignment between offered and requested data.
\end{abstract}


\IEEEpeerreviewmaketitle

\section{Introduction}

\IEEEPARstart {L}{ow} Earth Orbit (LEO) satellite communication systems, which play a crucial role in 6G, have advanced quickly due to their wide coverage, low latency, high transmission capacity, and flexible resource scheduling \cite{9582617, 10143230, 10001098, 10437178}. To provide broadband transmission and seamless coverage, LEO constellations must ensure multiple coverage, where some areas are served by multiple satellites. However, the uneven geographic distribution of ground users and the high mobility of LEO satellites lead to temporal and spatial variability in traffic demands within a satellite's coverage area. Beam Hopping (BH) technology, based on time slicing, achieves the traditional multi-beam coverage with fewer beams. This makes LEO satellites equipped with BH antennas highly flexible, which is crucial for addressing the varying demands across different regions \cite{9955995}. Nonetheless, if beams are powered uniformly, the transmission capacity will still fall short of meeting the uneven traffic demands across different beams \cite{1545873, 8333695, 6112304}. Additionally, onboard power in LEO satellites is extremely limited \cite{8700141}, making the intelligent and real-time joint BH and power allocation (PA) a critical research focus in LEO satellite communication systems.

Nevertheless, there are still some critical issues which are closely related to practical implementations and have not been well addressed in effectively scheduling beams and power resources in LEO satellite constellations \cite{1545873, 6112304}. First, the constantly changing topology of LEO satellite networks leads to fluctuations in inter-beam interference and satellite-to-ground link conditions, demanding adaptive rather than static resource management \cite{10286908, 10292611}. Second, managing resource scheduling is key to reducing beam interference, both within and between satellites. Third, joint optimization of beam and power resources expands the system's state-action space, exacerbating the "curse of dimensionality." Additionally, this joint optimization creates a discrete-continuous hybrid action space, where beam allocation is discrete, and PA is continuous \cite{9439849}. Approximating continuous actions using finite discrete sets diminishes the natural structure of continuous actions, while relaxing discrete actions into continuous sets significantly complicates the action space. Finally, scheduling beams and power resources must not be viewed as a simple single-objective optimization, but should concurrently consider multiple performance metrics, such as network throughput, latency, and fairness.

In reviewing the literature, it is evident that most studies concentrated on flexible radio resource allocation with a fixed beam direction, applicable to either a single satellite \cite{8700251, 8976431, 1545873, 8642812} or multiple satellites \cite{9512414, 9460776, 9420293, 9791128}. Some studies (e.g., \cite{9502162, 9928761}) primarily utilize BH technology to address the uneven geographical distribution of users. However, these studies overlook the importance of adaptive beam resource allocation, opting instead for a static allocation strategy, such as equal distribution across beams. Although certain works \cite{9479792, 9769901, 8957062} are dedicated to joint optimization of PA and beam direction, they are limited to single-satellite scenarios. 

In summary, existing research cannot be directly applied to multi-beam, multi-LEO satellite networks to address the aforementioned key issues. Furthermore, addressing all of these issues simultaneously presents new challenges for the dynamic resource management in LEO satellite networks due to the following aspects. 
\begin{enumerate}[label=\textbf{\alph*.}]
\item Inter-beam interference is unavoidable when multi-beam LEO satellites transmit downlink data through BH. This interference includes both intra-satellite and inter-satellite interference, leading to a strong interdependence between resource allocation decisions across satellites. Therefore, to fully exploit the spectrum reuse benefits in multi-beam LEO satellite communications, fostering cooperation among satellites is crucial to mitigate interference and enhance system capacity. However, the limited computational capacity and energy resources of satellites significantly increase the difficulty of satellite cooperation.

\item The resource allocation decisions (including BH and PA) should be jointly optimized to maximize overall network performance. Since BH and PA are closely intertwined, their joint optimization is typically NP-hard, making it difficult to achieve an optimal solution within polynomial time.

\item With the constant movement of LEO satellites, the connections between satellites and ground terminals frequently change. Consequently, resource allocation decisions must be adaptable to this dynamic environment, further increasing the complexity of resource management.
\end{enumerate}

To address these complex challenges, we note that Digital Twin (DT) technology provides an opportunity to represent the real world in a virtualized manner and has emerged as a promising tool for guiding resource deployment and scheduling in the real world. As an effective method for bridging the gap between physical entities and the digital realm, DT has garnered significant attention from both industry and academia \cite{9905999}. In industries such as manufacturing, power grids, and transportation management, DT has proven its ability to enhance operational efficiency and reduce resource waste through precise system modeling and predictive capabilities \cite{10034656}. In multi-LEO satellite communication systems, introducing DT can provide a global perspective for LEO satellite systems. By virtually replicating and monitoring the status and resource dynamics of physical satellite networks, DT can fully capture the real-time status changes of satellites, ground users, and the environment \cite{10246262}. This global perspective can support the overall optimization of BH and PA, not only helping to coordinate resource allocation for multiple satellites, but also effectively dealing with interference management and load balancing issues between satellites.

Motivated by these considerations, we design a DT-empowered multi-beam, multi-satellite cooperative service network and formulate a resource management problem for the downlink scenario that integrates both BH and PA. Our objective is to enhance the fairness of cell service while simultaneously improving throughput. To address this esource management problem, we decompose the original problem into two sub-problems: demand prediction-based BH at the DT layer and beam PA at the LEO satellite layer. We propose two advanced reinforcement learning algorithms to solve these sub-problems. The key contributions of this paper are summarized as follows:

\begin{itemize}
\item We propose a DT-empowered resource allocation framework for LEO satellite networks, addressing the joint BH and PA problem in a dynamic multi-beam, multi-satellite downlink scenario. The objective is to maximize throughput while ensuring fairness across cells. Given the mixed-integer, non-convex nature of the optimization problem, we decompose it into two simpler sub-problems: a DT layer multi-LEO satellite cooperative BH problem focused on load balancing, and a multi-beam power resource allocation problem at the LEO layer.

\item For optimization at the DT layer, the cloud predicts future time slot requirements based on historical demand data, considering the overlapping coverage of LEO satellite systems to design a BH scheme that balances communication demands among satellites. We employ the A3C algorithm, deploying an actor-critic network on each satellite and a global critic network in the cloud to improve training efficiency and stability.

\item To optimize resource allocation at the LEO layer, each satellite beam is modeled as an agent, and an on-demand beam PA problem is established. The Multi-Agent Deep Deterministic Policy Gradient (MADDPG) algorithm is utilized to facilitate cooperation and competition among these agents, further enhancing cell service fairness and system throughput.

\item Extensive simulation results demonstrate that the proposed joint BH and PA algorithm significantly outperforms baseline algorithms in terms of throughput, load balancing, and cell fairness.
\end{itemize}

The rest of this paper is organized as follows. Section II reviews the related literature. Section III describes the system model and presents the joint BH and power resource allocation optimization problem for multiple LEO satellites. Section IV decomposes the optimization problem into two sub-problems and provides a detailed explanation of the solution methods for these sub-problems. Section V presents the simulation results, demonstrating the superior performance of the proposed method. Section VI presents a summary of this paper and discusses future research prospects.

\section{Related Work}
Recently, to better align beam resources with non-uniform traffic demands, flexible resource allocation strategies for satellite communication systems have garnered significant attention \cite{8957062, 9443991, 8695098, 9417306, 10063711, 8766489, 9693289, doi:10.1177/1550147717709461, 9013589, 9558712, 9616240, 7039249, 10582841, 1545873, 9460776, 9420293, 10.1007/978-981-13-5937-8_13, 8971890, 10495871, 10124063, 9615707, 9968247, 9916289}. 

For instance, \cite{1545873} presented a PA and multi-beam scheduling method to optimize limited power resources against unevenly distributed user traffic, considering the trade-off between total capacity and user fairness. \cite{9460776} focused on subchannel and PA optimization in multi-beam GEO satellite communication systems, aiming to satisfy traffic demands with minimal transmission power and bandwidth. In \cite{9420293}, the authors formulated a joint optimization problem of PA and data transmission scheduling for satellite-assisted remote IoT networks, with the objective of maximizing total data rate, and they designed a model-free reinforcement learning framework to accommodate the highly dynamic nature of LEO satellite communications. 

Similarly,  Takahashi et al. \cite{8695098} introduced a novel mathematical model that balances beam pointing and PA, significantly improving satellite communication resource utilization and highlighting the benefits of joint optimization over single-dimensional optimization. Wang et al. \cite{9417306} proposed a joint optimization algorithm for PA, beam scheduling, and terminal time-slot allocation in coexisting BH-NOMA systems, devising a constraint scheme to accurately gauge the global optimal solution. Du et al. \cite{10063711} proposed a dynamic reinforcement learning-based strategy to integrate beam hopping and power allocation (BHPA), enhancing power utilization and satellite throughput while demonstrating the feasibility of PA. 

Further,  Shi et al. \cite{doi:10.1177/1550147717709461} addressed the multi-objective optimization problem of system throughput and terminal delay, employing a heuristic method to jointly allocate power and bandwidth. The authors in \cite{7039249} developed a two-stage approach for beam PA, utilizing genetic algorithm and simulated annealing (GA-SA). Lin et al. \cite{10582841} presented a two-stage BH design strategy aimed at minimizing energy consumption in BH satellite communications while addressing users' heterogeneous demands.

However, the aforementioned studies primarily focus on single GEO satellites. Joint BH and resource allocation for multiple LEO satellites presents distinct challenges. The rapid motion of LEO satellites, coupled with constantly changing channel conditions and varying traffic demands, differentiates the LEO scenario from GEO systems. Additionally, LEO terminals are typically small and prone to interference from neighboring satellites, especially in regions covered by multiple satellites. This makes it challenging to directly apply GEO-based BH resource allocation schemes to LEO systems.

In response, recent research has begun investigating resource allocation strategies for multi-beam LEO satellite systems\cite{10.1007/978-981-13-5937-8_13, 8971890, 10495871, 10124063, 9615707, 9968247, 9916289}. Liu et al. \cite{10.1007/978-981-13-5937-8_13} modeled the dynamic coverage area of LEO satellites as rectangular blocks and applied an iterative algorithm to optimize system capacity. A greedy algorithm based on beam position traffic demand was utilized to enhance throughput \cite{8971890}. Shuang et al.\cite{10495871} introduced a joint beam scheduling and power optimization algorithm considering the geographic distribution of aggregation nodes, while examining the impact of co-channel interference in LEO systems, focusing primarily on single LEO satellites. Li \cite{10124063} presented an optimization solution for the high-dimensional problem in dual-satellite scenarios. Deng \cite{9615707} investigated system capacity optimization in three-layer heterogeneous satellite networks, focusing on load balancing and interference management. Lin et al. \cite{9968247, 9916289} explored the BH scheme for NGSO multi-satellite systems and the coexistence of NGSO and GSO satellites. These studies, however, assume even power distribution across satellite beams, leaving opportunities for further optimization.

The most closely related work is \cite{10398511}, which explored the joint optimization of beam direction alongside spectrum, time, and power resources in dmicyna multi-beam LEO satellite networks. Nonetheless, the overlapping beam coverage of multiple LEO satellites introduces the risk of interference between adjacent satellites. Although beam direction control has been addressed, the issue of interference management in multi-satellite cooperation has not been fully resolved. This paper aims to tackle this problem comprehensively.

\section{System Model And Problem Formulation}
This section first describes the system model, which consists of the digital twin network, traffic model, and communication model. Next, the joint optimization problem of BH and PA is formulated. For convenience, the detailed notations and definitions used in this paper are summarized in Table I.

\begin{table}[h!]
\centering
\caption{Notations and Definitions}
\label{tab:notations}
\begin{tabular}{l|l}
\hline
\text{Notations} & \text{Definitions} \\ \hline
$N$ & Number of satellites \\ 
$C$ & Number of cells covered per satellite \\ 
$K$ & Number of beams per satellite\\
$B$ & Total satellite bandwidth \\ 
$T$ & BH period \\ 
$J$ & Delay fairness \\ 
$Q$ & Load balancing \\ 
$\alpha$ & The weight between Load balancing and delay fairness \\ 
$\beta$ & The weight between throughput and delay fairness \\
$V_{n}$ & The set of cells under the coverage of satellite $n$\\
$T_{ttl}$ & Time to live of a data packet in queue \\ 
$\rho_{n,c}^t$ & Arrival traffic in time slot $t$ for cell $c$ of satellite $n$\\
$\hat{\rho}_{n,c}^t$ & Predicted of arrival traffic at slot $t$ for cell $c$ of satellite $n$\\
$\lambda_n^t$ & The arrival rate of cell $c$ covered by satellite $n$ in slot $t$ \\ 
$d_{n,c}^t$ & The total traffic stored in queue $c$ at slot $t$ in satellite $n$ \\ 
$\phi_{c,l}^{n,t}$ & Data packets stored in queue $c$ for $l$ time slots \\ 
$x_{n,c}^t$ & The illumination status of cell $c$ by satellite $n$ in time slot $t$ \\
$L_n^t$ & The traffic load of satellite $n$ at slot $t$\\ 
$\tau_{n,c}^{t}$ & Average queue delay per data packet in cell $c$ at time slot $t$ \\ 
$T_{rx}$ & The noise temperature of the receiver\\ 
$P_{tot}$ & The total power of each satellite \\
$p_{n,k}^t$ & Power of beam $k$ belonging to satellite $n$ \\ 
$P_{max}$ & The maximum power allocated to each beam \\ 
$h_{k_c,l}$ & Channel coefficients from beam $l$ to user $k_c$ in cell $c$. \\
$G_{t}(\theta_{k_c, l})$ & Transmit antenna gain from the
$l$-th beam to the $k_c$-th user \\
$G_{r}^{k_c}$ & Receive antenna gain of the user $k_c$ \\
$r_{n,c}^t$ & Channel capacity of cell $c$ served by satellite $n$ in slot $t$ \\ 
$\omega_{i, j}$  & The distance between cells $i$ and $j$ \\
$\varpi$  & The minimum interference distance between cells\\ 
$Th_{n,c}^t$ & The throughput of cell $c$ during time slot $t$ \\\hline
\end{tabular}
\end{table}

\subsection{System Model}
\subsubsection{Digital Twin Network} 
This paper considers the forward link of a multi-beam LEO satellite communication system in a DT-empowered satellite network, consisting of ground users, LEO satellites, and a cloud center. As shown in Fig. 1, LEO satellites serve users in ground coverage cells through BH. Adjacent LEO satellites have overlapping coverage areas, meaning that multiple satellites may cover the same cell at the same time.

\begin{figure}[htbp]
\centering
\includegraphics[width=8cm]{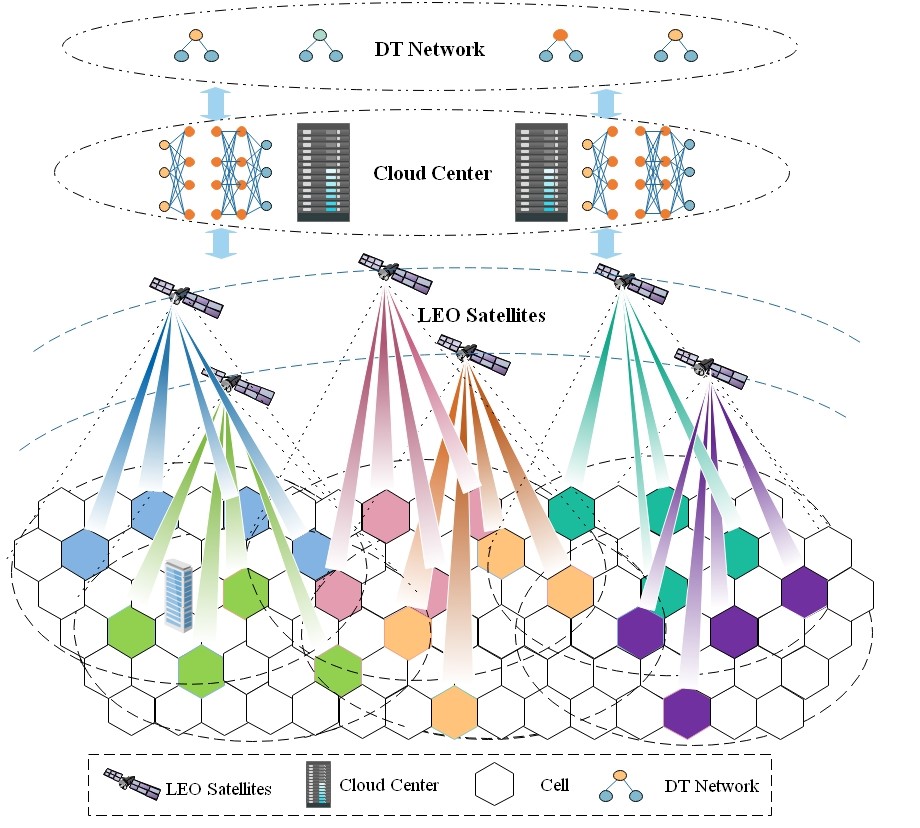}
\caption{Forward link of the DT-empowered multi-beam LEO satellite communication system.}
\vspace{-0.5cm}
\label{fig1}
\end{figure}

Adjacent LEO satellites have overlapping coverage areas, meaning multiple satellites may cover a cell. The cloud center maintains a DT model of the entire network to monitor and optimize resource allocation. It periodically collects global network state information, such as cell requests, LEO satellite resource usage, and buffer queue backlogs. Using this DT, the cloud center monitors dynamic changes in the LEO satellite network and optimizes BH plans. The DT network determines the service cells for each time slot, encompassing the allocation of overlapping cells.

Consider that there are $N$ satellites serving the ground area containing a total of $M$ cells. The sets of satellites and cells are denoted by $\mathcal{N} = \{n|n=1,2,\cdots, N\}$, and $\mathcal{V} = \{v|v=1,2,\cdots, V\}$, respectively. Each satellite can produce $K$ beams and provide transmission service to $C$ cells through time-division multiplexing (TDM) manner. The set of $K$ beams of satellite $n$ is expressed as $\mathcal{K}_n = \{k_n|k_n=1,\cdots,k,\cdots, K\}$. 
In particular, the set of cells covered by $n$-th satellite is represented by $V_n$,  $|V_n|= C$. Owing to the multiple coverage characteristics of the LEO satellite communication system \cite{9916289}, one cell may be covered by multiple satellites, i.e. $V_i \cap V_{j} \neq \emptyset$, $\forall i, j\in \mathcal{N}$. Without loss of generality, we consider the first $T$ time slots, denoted as $\mathcal{T} = \{t | t = 1, \cdots,t,\cdots, T\}$. The duration of each time slot is $T_{slot}$.  

\subsubsection{Traffic Model} 
Assuming that each satellite holds $C$ queues for storing the arriving traffic from the coverage cells, and each queue capable of storing only the traffic that arrived within the last $T_{ttl}$ time slots due to the limited queue storage, where $T_{ttl}$ is defined as Time To Live (TTL) \cite{8971890}. $\boldsymbol{\Lambda}^t_n = \{\rho_{n,1}^t,\cdots,\rho_{n,c}^t,\cdots,\rho_{n, C}^t\}$ denotes the arrival traffic of satellite $n$ in time slot $t$, where $\rho_{n,c}^t$ represents the arrival traffic of cell $c$ covered by satellite $n$ in time slot $t$. Therefore, the total traffic stored in satellite $n$ during time slot $t$ is represented as $\boldsymbol{D}^{t}_n = \{d_{n,1}^{t},\cdots,d_{n,c}^{t},\cdots,d_{n, C}^{t}\}$, where $d_{n,c}^{t}$ is the traffic demand of cell $c$ in time slot $t$. We can reasonably expand $d_{n,c}^{t}$ to $\{\phi_{c, 1}^{n,t},\phi_{c, 2}^{n,t}, \cdots, \phi_{c, T_{ttl}}^{n,t}\}$, where $\sum_{l=1}^{T_{ttl}} \phi_{c,l}^{n,t}=d_{n,c}^{t}$ and $\phi_{c,l}^{n,t}$ represents the number of data packets that have been in queue $c$ for $l$ time slots. For users in overlapping coverage cells, all covering satellites will receive their requests.

To satisfy the traffic demand of cells, each satellite should reasonably decide the beam irradiation positions. The BH pattern of satellite $n$ at time slot $t$ can be expressed as
\begin{equation}
\boldsymbol{X}_n^t=\{{x}_{n,1}^t, \ldots, {x}_{n,c}^t, \ldots, {x}_{n,C}^t\}
\end{equation}
where ${x}_{n,c}^t \in \{0,1\}$ denotes whether the coverage cell $c$ of satellite $n$ is illuminated by the beam of satellite $n$ in time slot $t$.

\subsubsection{Communication Model}
In order to fully utilize spectrum resources and enhance spectrum efficiency, all beams are considered to occupy the entire satellite bandwidth $B$. This implies the use of full frequency reuse in the multi-satellite BH system, where co-frequency interference between beams becomes significant. For example, a user in the cell $c$ served by the $k$-th beam (referred to as the $k_c$-th user) will receive $NK$ signals from $N$ satellites. The channel vector linking the user $k_c$ to $NK$ beams is expressed as $\boldsymbol{h}_{k_c}=\left[{h}_{k_c,1}, {h}_{k_c,2}, \ldots, {h}_{k_c,NK}\right]$. Assuming the user compensates for the Doppler frequency offset due to satellite motion and that clear sky conditions prevail, neglecting rain attenuation.
Let $G_t\left(\theta_{k_c, l}\right)$ denote the transmit antenna gain of the $l$-th beam to user $k_c$, which is determined by the off-axis angle $\theta_{k_c, l}$ between user $k_c$ and the main lobe direction of beam $l$. The receiving antenna gain for user $k_c$ is denoted by $G_r^{k_c}$. So, the channel coefficient between the $k_c$-th user and the $l$-th beam can be denoted by \cite{10582841},
\begin{equation}
h_{k_c, l}=\frac{\sqrt{G_t\left(\theta_{k_c, l}\right) G_r^{k_c}}}{4 \pi \frac{d_{k_c, l}}{\lambda^{'}}}
\end{equation}
where $d_{k_c, l}$ is the distance between user $k_c$ and beam $l$, and $\lambda^{'}$ represents the wavelength.

Therefore, the signal-to-noise ratio of user $k_c$ can be expressed as,
\begin{equation}
\begin{aligned}
\text {SINR}^{t}_{n,k_c}= \frac{p_{n,k}^t\left|h_{{k_c}, k}\right|^2}{\sigma ^2+I_{Intra}+I_{Inter}}
\end{aligned}
\end{equation}
where $\sigma ^2=k_B T_{r x}B$ represents the noise power. $k_B$ is the Boltzmann constant and $T_{r x}$ is the receiver noise temperature. $I_{Intra}=\sum_{l \in \mathcal{K}_n \backslash\{k\}} p_{n,l}^t \left|h_{{k_c}, l}\right|^2$ denotes intra-satellite interference, $I_{Inter}=\sum_{n^{\prime} \in \mathcal{N} \backslash\{n\}} \sum_{l \in \mathcal{K}_{n^{\prime}}} p_{n^{\prime},l}^t \left|h_{{k_c}, l}\right|^2$ represents inter-satellite interference. $p_{n,l}^t$ is the power allocated by satellite $n$ to its $l$-th beam. Similarly, $p_{n^{\prime},l}^t$ is the power allocated by satellite $n^{\prime}$ to its $l$-th beam.

According to (3), let $r_{n,c}^{t}$ denote the channel capacity of cell $c$ served by satellite $n$ in time slot $t$ which can be calculated as follows:
\begin{equation}
r_{n,c}^{t}=x_{n, c}^tB \log _2\left(1+SINR_{n,k_c}^{t}\right)
\end{equation}
where $x_{n, c}^t$ denotes if the cell $c$ is illuminated, and $B$ denotes the bandwidth allocated to beam $k$ serving cell $c$. Since the system employs full-frequency multiplexing, each beam utilizes the entire available bandwidth $B$.

Let $Th_{n,c}^{t}$ denote throughput of cell $c$ at time slot $t$, which can be calculated as follows:
\begin{equation}
Th_{n,c}^{t}=\min \left\{r_{n,c}^{t}, d_{n,c}^{t}\right\}
\end{equation}

\subsection{Optimization Problem Formulation}
It should be noted that multiple beams from different satellites may serve the same cell at the same time slot, resulting in significant inter-satellite interference. Therefore, it is necessary to carefully plan the BH pattern of each satellite and the corresponding PA of the beams. Given the time-varying characteristics of traffic demands, the joint BH and PA aims to maximize the total throughput of the satellite while minimizing delay fairness among all cells, given the limitations of total bandwidth and power resources.

The total throughput of all satellites can be expressed as follows:
\begin{equation}
Th_{tot} = \sum_{t=1}^{T}\sum_{n=1}^{N}\sum_{c=1}^{C} Th_{n,c}^{t}
\end{equation}

Likewise, let $\tau_{n,c}^{t}$ denote the average queueing delay of each data packet of the cell $c$ during time slot $t$, which can be computed by
\begin{equation}
\tau_{n,c}^{t}=\frac{\sum_{l=1}^{T_{t t l}} l \cdot \phi_{c, l}^{n,t}}{\sum_{l=1}^{T_{t t l}} \phi_{c, l}^{n,t}}
\end{equation}

Therefore, the delay fairness among all cells $J$ is expressed as follows:
\begin{equation}
J=\sum_{n=1}^{N}  \sum_{t=1}^T \left(\max _{c \in {V_{n}}}\left\{\tau_{n,c}^{t}\right\}-\min _{c \in {V_{n}}}\left\{\tau_{n,c}^{t}\right\}\right)
\end{equation}

To maximize throughput while reducing latency fairness, we establish a optimization problem. This problem combines BH and PA, and can be modeled as follows:
\begin{equation}
\begin{aligned}
{P_0:}&\max _{\boldsymbol{X}^t_{n},  \boldsymbol{P}^t_n} {\beta \frac{Th_{tot }}{T h_{norm }}-(1-\beta) \frac{J}{J_{norm }}}, \beta \in[0,1]\\
\text { s.t. } & C_1: x_{n,c}^t \in \{0,1\},  \forall n \in \mathcal{N}, \forall c \in V_n  \\
\vspace{5pt}
& C_2: \sum_{c=1}^C x_{n,c}^t=K, \forall n \in \mathcal{N} \\
\vspace{5pt}
& C_3: \sum_n x_{n, c}^t \leq 1,  \forall n \in \mathcal{N}\\
\vspace{5pt}
& C_4: x_{n, c}^t=0, \forall c \notin V_n , \forall n \in \mathcal{N}\\
\vspace{5pt}
& C_5: x_{n,i}^t x_{n',j}^t=1, \omega_{i, j} \geq \varpi, \forall i \in V_n, j \in V_{n'}\\
\vspace{5pt}
& 
C_6: p_{n,k}^t\leq P_{max}, \forall n \in \mathcal{N}, \forall k \in \mathcal{K}_n\\
\vspace{5pt}
& C_7: \sum_{k=1}^K p_{n,k}^t \leq P_{tot},  \forall n \in \mathcal{N}
\end{aligned}
\end{equation}

In the aforementioned optimization problem, $\boldsymbol{X}_{n}^t$ and $\boldsymbol{P}^t_{n}$ are the variables to be optimized, representing the BH pattern and the corresponding beam PA of satellite $n$ at time slot $t$, respectively. The parameter $\beta$ is a preset weight used to balance the trade-off between throughput and delay fairness, while $Th_{max}$ and $J_{max}$ are normalization constants. $C_1$ constrains optimization variables $x_{n,c}^t$ to be binary variables. $C_2$ states that each satellite can activate exactly $K$ beams at once, meaning each satellite can select $K$ cells within its coverage area for communication services in each time slot. $C_3$ emphasizes that each cell is served by at most one beam from one satellite in each time slot. $C_4$ means that each satellite can only select cells within its coverage area for service. $C_5$ enforces that the distance between cells $i$ and $j$ serviced in time slot $t$ exceeds the minimum interference distance $\varpi$, with $\omega_{i, j}=\operatorname{dist}(i, j)$ denoting the distance between cells $i$ and $j$. $C_6$ limits the power $p_{n,k}^t$ allocated to each beam not exceeding the maximum beam power $P_{max}$. $C_7$ ensures that the total beam power of satellite $n$ does not exceed the total power of satellite $P_{tot}$.

It can be observed that problem (9) is both non-convex and nonlinear, with the non-convexity arising from the SINR in equation (3). Additionally, problem (9) involves continuous variables $p_{n,k}^t$ and discrete variables $x_{n,c}^t$. As a result, this optimization problem is NP-hard, making it impractical to find an optimal solution in polynomial time. In the following section, we will introduce a new method that provides a sub-optimal solution with low computational complexity.

\section{Problem Decomposition and Proposed BHPA-LBDP Algorithm}
In this section, $P_0$ is divided into two sub-problems: the BH design problem for multiple satellites based on load balancing and interference avoidance, and the PA problem for each satellite.

\subsection{Multi-Satellite BH Design Based on Load Balancing and Interference Avoidance }\label{AA}
In LEO satellite BH systems, the distribution of ground services is uneven, resulting in significant differences in service demands across satellites. Certain satellites cover high-demand, densely populated regions, whereas others cover sparsely populated areas with relatively low demand. If all high-demand cells were assigned to a single satellite, it would become overloaded, leaving the resources of other satellites underutilized. As a result, balancing the load among satellites is crucial for improving the system's overall resource efficiency.

Since hotspot areas can be covered by multiple neighboring satellites simultaneously, serving the same hotspot cell with multiple satellites at the same time can result in severe inter-satellite interference. Therefore, assigning these hotspot areas to different satellites can reduce the load on individual satellites and avoid interference among neighboring satellites.

\subsubsection{Optimization Problem Formulation} \label{BB}
In summary, the multi-satellite traffic-driven BH problem can be described as minimizing the weighted sum of traffic load between satellites and cell service fairness, which is formalized as follow:
\begin{equation}
\begin{aligned}
P_1:& \underset{\boldsymbol{X}^t_{n}}{\operatorname{min}}\left\{\alpha\frac{Q}{Q_{max }}+(1-\alpha) \frac{J}{J_{max }}\right\} \\
 \text { s.t. } & C_1: Q=\sum_{t=1}^T\max _{n \in \mathcal{N}}\left\{L_{n}^{t}\right\}-\sum_{t=1}^T\min _{n \in \mathcal{N}}\left\{L_{n}^{t}\right\} \\
&  C_2: L_n^{t}=\sum_{c\in V_n} d_{n,c}^t x_{n, c}^t, \forall n \in \mathcal{N}\\
&  C_3: C_1-C_5  \quad \text {from}  \quad P_0
\end{aligned}
\end{equation}
where $\boldsymbol{X}_n^t$ is the only optimization variable of the optimization problem, representing the BH pattern of satellite $n$ at time slot $t$. In $C_1$, $Q$ represents the load balance among satellites, defined as the difference between the maximum and minimum satellite loads. $L_n^{t}$ represents the traffic load of satellite $n$, which is calculated in $C_2$.

It should be noted that the optimization problem (10) is an integer, non-linear, and non-convex problem, which is still NP-hard. Additionally, due to the rapid movement of LEO satellites, the coverage cells of the satellites are dynamically changing, that is, the arrival traffic of the coverage cells received by the satellites is frequently changing. Thus, traditional integer optimization methods, such as branch and bound, cannot be applied here. In this paper, an DRL algorithm is proposed to find the relatively optimal BH decision through continuously interacting with the environment.

\subsubsection{Multi-Agent MDP Formulation} \label{CC} In fact, the BH optimization problem (10) is essentially a sequential decision-making problem, and the traffic $\boldsymbol{D}^t$ can be expressed as,
\begin{equation}
\boldsymbol{D}^t=\boldsymbol{D}^{t-1}-\boldsymbol{X}^{t-1} \boldsymbol{r}^{t-1} + \boldsymbol{\Lambda}^t
\end{equation}

However, the request arrival traffic state $\boldsymbol{\Lambda}^t$ is unknown during BH decision-making. Fortunately, the states from previous time slots are easily accessible. Given the temporal correlation between state changes, we are motivated to predict the current state by leveraging past state sequences.

\textbf {Arrival Traffic Estimation:} The LSTM model is trained using historical request arrival traffic state sequences of cells, which are collected by DT. The error between the actual state and the predicted state is used to update the weight parameters $W$ and $U$. At time step $t-1$, the request arrival traffic state $\rho_{n,c}^{t-1}$ serves as the input to the LSTM network, with the output providing an estimate of the current request arrival traffic state. The 
memory block parameters are updated as follows:
\begin{equation}
\label{eq12}
\begin{split}
\boldsymbol{f}_t&=\sigma\left[\boldsymbol{W}_f \boldsymbol{\rho}_{n,c}^{t-1}+\boldsymbol{U}_f \boldsymbol{h}_{t-1}+\boldsymbol{b}_f\right]\\
\boldsymbol{i}_t&=\sigma\left[\boldsymbol{W}_i \boldsymbol{\rho}_{n,c}^{t-1}+\boldsymbol{U}_i \boldsymbol{h}_{t-1}+\boldsymbol{b}_i\right]\\
\hat{\boldsymbol{C}}_t&=\tanh \left[\boldsymbol{W}_c \boldsymbol{\rho}_{n,c}^{t-1}+\boldsymbol{U}_c \boldsymbol{h}_{t-1}+\boldsymbol{b}_c\right]\\
\boldsymbol{C}_t&=\boldsymbol{f}_t \odot \boldsymbol{C}_{t-1}+\boldsymbol{i}_t \odot \widetilde{\boldsymbol{C}}_t\\
\boldsymbol{o}_t&=\sigma\left[\boldsymbol{W}_o \boldsymbol{\rho}_{n,c}^{t-1}+\boldsymbol{U}_o \boldsymbol{h}_{t-1}+\boldsymbol{b}_o\right]\\
\boldsymbol{h}_t&=\boldsymbol{o}_t \odot \tanh \left(\boldsymbol{C}_t\right)\\
\hat{\boldsymbol{\rho}}_{n,c}^{t}&=\boldsymbol{W}_h \boldsymbol{h}_t+\boldsymbol{b}_h
\end{split}
\end{equation}
Here, $\boldsymbol{W}_f$ represents the weight between the input and the forget gate, while $\boldsymbol{U}_f$ corresponds to the weight between the previous hidden state $\boldsymbol{h}_{t-1}$ and the forget gate. The bias is $\boldsymbol{b}_f$, and $\sigma(\cdot)$ denotes logical sigmoid function. $\boldsymbol{W}_i$ and $\boldsymbol{W}_c$ are the weights linking the input state $\boldsymbol{\rho}_{n,c}^{t-1}$ to the input gates, whereas $\boldsymbol{U}_i$ and $\boldsymbol{U}_c$ represent the weights between the previous hidden state $\boldsymbol{h}{t-1}$ and the input gate.  The hyperbolic tangent function is $\tanh(\cdot)$, and $\odot$ stands for the Hadamard product. $\boldsymbol{W}_o$ refers to the weight between the current input $\boldsymbol{\rho}_{n,c}^{t-1}$ and the output gate, and $\boldsymbol{U}_f$ represents the weight between the hidden state $\boldsymbol{h}_{t-1}$ and the output gate. Finally, $\boldsymbol{W}_h$ and $\boldsymbol{b}_h$ are the output weight matrix and bias term corresponding to the predicted state $\hat{\boldsymbol{\rho}}_{n,c}^{t}$, respectively.

Thus, we can model the state evolution using a Markov Decision Process (MDP). Specifically, the four key components of the MDP, i.e., state, observation, action, and reward, are defined as follows.
\begin{itemize}
\item \textbf {State:} At time slot $t$, the system can observe the entire traffic demand as the global state, which is represented as follows:
\begin{equation}
\label{eq13}
\begin{split}
\boldsymbol{s}^t&=[\operatorname{vec} (\mathbf{\hat D}^t); \operatorname{vec} (\mathbf{H}^t)]\\
\boldsymbol{\hat D}^t&=\boldsymbol{D}^{t-1}-\boldsymbol{X}^{t-1} \boldsymbol{r}^{t-1} + \boldsymbol{\hat \Lambda}^t\\
\boldsymbol{D}^{t-1}&=\left[\begin{array}{cccc}
d_{1, 1}^{t-1} & d_{1, 2}^{t-1} & \cdots & d_{1, C}^{t-1} \\
d_{2, 1}^{t-1} & d_{2, 2}^{t-1} & \cdots & d_{2, C}^{t-1} \\
\cdots & \cdots & \cdots & \cdots \\
d_{N, 1}^{t-1} & d_{N, 2}^{t-1} & \cdots & d_{N, C}^{t-1}
\end{array}\right]\\
\boldsymbol{\hat\Lambda}^t&=\left[\begin{array}{cccc}
\hat\rho_{1, 1}^{t} & \hat\rho_{1, 2}^{t} & \cdots & \hat\rho_{1, C}^{t} \\
\hat\rho_{2, 1}^{t} & \hat\rho_{2, 2}^{t} & \cdots & \hat\rho_{2, C}^{t} \\
\cdots & \cdots & \cdots & \cdots \\
\hat\rho_{N, 1}^{t} & \hat\rho_{N, 2}^{t} & \cdots & \hat\rho_{N, C}^{t}
\end{array}\right]\\
\boldsymbol{H}^t&=\left[\begin{array}{cccc}
h_{1, 1}^{t} & h_{1, 2}^{t} & \cdots & h_{1, C}^{t} \\
h_{2, 1}^{t} & h_{2, 2}^{t} & \cdots & h_{2, C}^{t} \\
\cdots & \cdots & \cdots & \cdots \\
h_{N, 1}^{t} & h_{N, 2}^{t} & \cdots & h_{N, C}^{t}
\end{array}\right]\\
\end{split}
\end{equation}

where $\hat\rho_{n, c}^{t}$ represents the predicted request arrival traffic in cell $c$ covered by satellite $n$ at time slot $t$, which can be obtained by (12). $h_{n, k}^{t}$ is channel gain from satellite $n$ to its $c$-th cell at time slot $t$, which can be obtained by (2). It should be noted that the state represents global information, which the agents (i.e., the satellites) cannot fully access since they only have local observations. Therefore, the global state can only be utilized by the centralized trainer in DT network. 

\item \textbf {Observation:} Denote $\boldsymbol{o}_n^t$ as the local observation accessed by the $n$-th agent at time slot $t$. Similar to the state, the local observation,
\begin{equation}
\label{eq14}
\begin{split}
\boldsymbol{o}_n^t=[\boldsymbol{D}_n^t;\boldsymbol{H}_n^t]
\end{split}
\end{equation}
where $\boldsymbol{D}_{n}^t=\left[d_{n, 1}^{t}, d_{n, 2}^{t} , \ldots , d_{n, C}^{t} \right]$ represents the status of traffic queues on the $n$-th satellite. $\boldsymbol{H}_{n}^t=\left[h_{n, 1}^{t}, h_{n, 2}^{t} , \ldots , h_{n, C}^{t} \right]$ is the channel information of the $n$-th satellite.

\item \textbf {Action:} As an agent, the satellite must decide which cells to illuminate. Thus, the action of agent can be defined as:
\begin{equation}
\begin{split}
\left.a_n^t\right|_{n \in N}&=\left(x_{n,1}^t,  \ldots, x_{n,c}^t, \ldots, x_{n,C}^t\right),\\ x_{n,c}^t&\in \{0,1\},  \text {and} \sum_{c=1}^C x_{n,c}^t=1
\end{split}
\end{equation}

\item \textbf {Reward:} To minimize the objective defined in equation (10), we design the reward at time $t$ as follows:
\begin{equation}
\begin{aligned}
&R^t=R^t\left(s^t, a^t\right)= -[\alpha \frac{\max \left\{ L_n^t \right\}-\min \left\{ L_n^t \right\}}{Q_{\text{max}}}+ \\
 &(1-\alpha) \frac{\max \left\{ \tau_{n, i}^t \right\}-\min \left\{ \tau_{n, j}^t \right\}}{J_{\text{max}}}+\Gamma], \forall i, j \in V_n
\end{aligned}
\end{equation}
Note that we introduce a penalty term $\Gamma$ in the reward to avoid inter-satellite interference. This penalty is proportional to the number of cells illuminated simultaneously. Thus, the reward will be reduced if multiple satellites illuminate the same cell at a time slot.
\end{itemize} 

The request arrival traffic estimation module deployed in DT can predict the state $\boldsymbol {\hat \rho}_{n}^{t}$ for time slot $t$ based on the observed state $\boldsymbol {\rho}_{n}^{t-1}$ for time slot $t-1$. The predicted state, along with the local queue state and channel state of time slot $t$ is then input into the local cooperative BH strategy module. This module outputs the local BH action $a$ that the agent should take at decision time $t$, thereby determining which cells will be served by satellite $n$.

\subsubsection{Multi-Agent BH Decision Based on A3C (MA3C-BH)} \label{DD}
Since the Actor-Critic method performs better in discrete action spaces, this paper employs a multi-agent A3C algorithm based on Actor-Critic method.  
\begin{figure}[htbp]
    \centering
    \includegraphics[width=8cm]{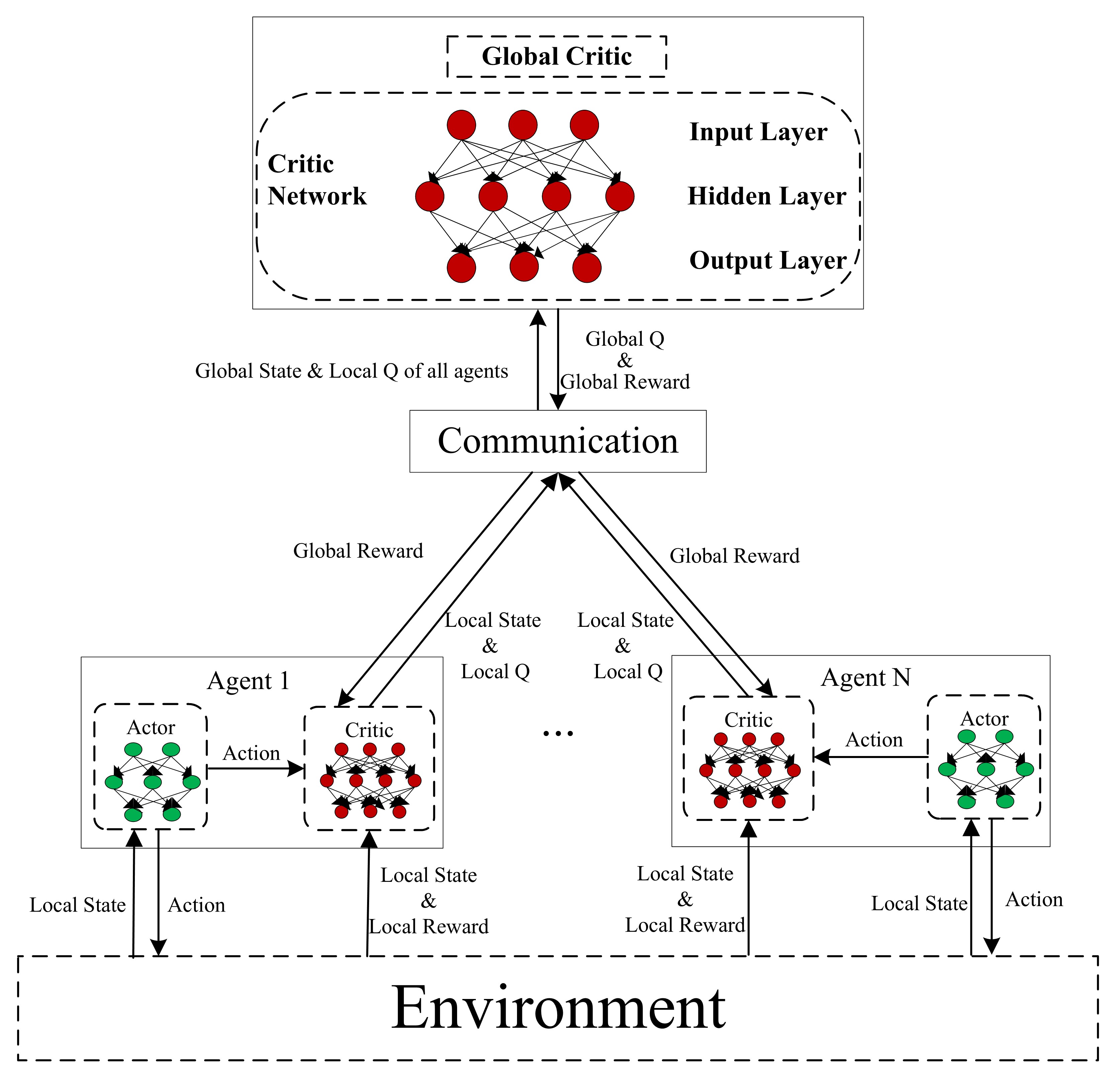}
    \caption{The cooperative multi-agent actor-critic framework of the MA3C-BH algorithm.}
    \label{fig2}
\end{figure}

Note that the BH decision is made in the DT, which has sufficient computing power. As shown in Fig.2, each satellite deploys an actor network to select actions and a critic network to estimate the value of actions. Simultaneously, the DT deploys a critic network to integrate the feedback of all agents and provide global value evaluation.

This study establishes $I+1$ threads based on the A3C algorithm. Each thread designs $N$ actor networks and $N+1$ critic networks as illustrated in Fig. 2, where $N$ represents the total number of satellites. The network parameters in each thread are denoted as $\{\theta_1,\theta_2,\cdots,\theta_N,\phi_1,\phi_2,\cdots,\phi_N\}$. The network $\{\theta_1,\theta_2,\cdots,\theta_N,\phi_1,\phi_2,\cdots,\phi_N\}$ outputs a decision network $\{a_1^t,a_2^t,\cdots,a_N^t\}$ based on the observation vector $\{\boldsymbol o_1^t,\boldsymbol o_2^t,\cdots,\boldsymbol o_N^t\}$ obtained from the environment. The Critic network $\{\phi_1,\phi_2,\cdots,\phi_N\}$ calculates the Temporal Difference-error (TD-error) $\delta$ based on the reward $r^t$ generated at that moment and passes it to the other actor networks. Each actor network uses the received TD-error $\delta$ to update its parameters via gradient descent.

\begin{algorithm}
\caption{Multi-agent A3C BH strategy (MA3C-BH)}
\begin{algorithmic}[1]
\State Initialize global parameters $\theta_g$ for actor network, $\phi_g$ for critic network
\State Initialize thread-specific parameters $\theta^i$ for actor network, $\phi^i$ for critic network
\State Initialize shared critic network parameters $\phi_s$
\State Set the number of threads $I$ and the number of actor-critic networks per thread $N$

\For{each thread $i = 1, \ldots, I$}
    \State Initialize environment
    \State Initialize local copy of parameters $\theta^i \leftarrow \theta_g$, $\phi^i \leftarrow \phi_g$
    \While{not converged}
        \State Receive state $s^t$ from environment
        
        \For{each actor-critic network $j = 1, \ldots, N$}
            \State Calculate action $a_j^t = \pi_{\theta_j^i}(s^t)$ using actor network
            \State Execute action $a_j^t$ and observe reward $r_j^t$ and next state $s^{t+1}_j$
            \State Calculate TD-error $\delta_j^i = r_j^t + \gamma V_{\phi_j^i}(s^{t+1}_j) - V_{\phi_j^i}(s^t)$
            \State Update actor network parameters $\theta_j^i$ using policy gradient by:
            $\Delta \theta_j^i=\nabla_{\theta_j^i} \log \pi_{\theta_j^i}\left(o_j^t, a_j^t\right) \delta_j^i $          
            \State Update local Critic network parameters $\phi_j^i$ using TD-error:
            $\phi_j^i \leftarrow \phi_j^i + \beta \delta_j^i \nabla_{\phi_j^i} V_{\phi_j^i}(s^t)$
        \EndFor
        
        \State Update shared critic network parameters $\phi_s$ using average TD error by (19)
    \EndWhile
\EndFor

\State Asynchronously update global parameters:
\[
\theta_g \leftarrow \frac{1}{I} \sum_{i=1}^{I} \theta^i
\]
\[
\phi_g \leftarrow \frac{1}{I} \sum_{i=1}^{I} \phi^i
\]
\end{algorithmic}
\end{algorithm}
In the A3C architecture, the global network does not participate in actual exploration. Instead, it copies and shares the parameters explored by the preceding threads, achieving asynchronous learning and accelerating convergence. In a separate thread, the critic network predicts the value function at two moments through the states $X(t)$ and $X(t+1)$, calculates the TD-error, and updates the network parameters by minimizing the least squares temporal difference (LSTD),
\begin{equation}
V^*=\arg \min _V\left(\delta_{\theta_c}^i\right)^2
\end{equation}
where $V^*$ represents the optimization value function. For the actor network, after receiving the TD-error calculated by the critic network, the TD-error is used to calculate its gradient:
\begin{equation}
\nabla_{\theta_j^i} J\left(\theta_j^i\right)=\mathbb{E}\left[\nabla_{\theta_j^i} \log \pi_{\theta_j^i}\left(o_j^t, a_j^t\right) \delta_j^i\right]
\end{equation}
where $\pi_{\theta_j^i}\left(s,a\right)$ represents the probability of the Actor network taking action $a_j^t$ under the state observation vector $o_j^t$. 

We then use gradient descent to update the actor network:
\begin{equation}
            \theta_j^i \leftarrow \theta_j^i + \alpha \nabla_{\theta_j^i} \log \pi_{\theta_j^i}\left(o_j^t, a_j^t\right) \delta_j^i
\end{equation}

Additionally, each agent will also use the shared critic network to update its value function:
\begin{equation}
\phi_s \leftarrow \phi_s + \lambda \frac{1}{V} \sum_{j=1}^{V} \delta_j^i \nabla_{\phi_s} V_{\phi_s}(s^t)
\end{equation}

In summary, the proposed algorithm MA3C-BH is showing in Algorithm 1.

\subsection{PA Based on Delay Fairness}
After solving problem $P_1$, the BH pattern is determined, and each satellite has to determine the PA. This scheme should promote the transmission of a larger proportion of packets in the satellite's buffer queue, preventing packet loss due to queuing delays exceeding the maximum tolerable threshold. This strategy is meticulously designed to enhance the satellite's data throughput while ensuring minimal delay fairness among the served cells, thus balancing efficiency and equity in resource allocation.

\subsubsection{Problem Formulation }
We define $p_{n,k}^t$ to represent the power allocated to the beam $k$ of satellite $n$ during slot $t$. From problem $P_0$ we can formulate the PA problem as:

\vspace{5pt}
$\begin{aligned}
P_2:& \max _{\boldsymbol{P}^t_n}\left\{\beta \frac{Th_{tot }}{T h_{max }}-(1-\beta) \frac{J}{J_{max }}\right\}, \beta \in[0,1] 
\end{aligned}$
\begin{equation}
\begin{array}{rr}
\vspace{5pt}
\text { s.t. }  C_1: C_6, C_7  \quad \text {from}  \quad P_0\\
\end{array}
\end{equation}
where $\boldsymbol{P}^t_{n}=\{{p}_{n,1}^t,\ldots, {p}_{n,k}^t,\ldots,{p}_{n,K}^t\}$ represents the power allocated to each beam of satellite $n$.
\subsubsection{The Sequential Decision-Making Problem Reformulation}

The problem $P_2$ can be regarded as a sequence decision-making problem, which can be further reformulated as an MDP with infinite states and continuous action spaces. Define the formulated MDP as $(\boldsymbol s^{t-1},\boldsymbol a^{t-1},\boldsymbol r^{t-1},\boldsymbol s^{t})$, where $\boldsymbol s$ represents the state set of the environment, $\boldsymbol a$ indicates the action taken by the agent, and $\boldsymbol r$ refers to the reward function.
\begin{itemize}
\item State: The states of systems consists of two parts: the matrix of the number of packets $\boldsymbol{\overline D}_n^t$ and the downlink loss matrix $\boldsymbol{\overline H}_n^{t}$. Thus, the state vector $\boldsymbol{s}^t$ can be defined as:
\begin{equation}
\label{eq22}
\begin{split}
\boldsymbol{s}^t&=(\boldsymbol {\overline D}_n^t,  \boldsymbol{\overline H}_n^{t}),\\
\boldsymbol{\overline D}_n^t&=\left[{d}_{n,c}^t |  {x}_{n,c}^{t}=1, \forall c \in V_n\right]\\
&=\left[{d}_{n,1}^t, \ldots, {d}_{n,k}^t, \ldots, {d}_{n,K}^{t}\right]\\
\boldsymbol{\overline H}_n^t&=\left[{h}_{n,c}^t |  {x}_{n,c}^{t}=1, \forall c \in V_n\right]\\
&=\left[{h}_{n,1}^t, \ldots, {h}_{n,k}^t, \ldots, {h}_{n,K}^{t}\right]\\
\end{split}
\end{equation}
where $\boldsymbol {\overline D}_n^t$ represents the traffic of $K$ cells selected to be served by satellite $n$ in time slot $t$, with $d_{n,k}^t$ and $h_{n,k}^t$ representing the traffic and channel gain of the selected $k$-th cell, respectively.
\item Action: In the LEO layer, each beam is considered as an agent that decides PA in each time slot. Thus, the actions can be defined as:
\begin{equation}
\boldsymbol{a}^t=\boldsymbol{a}_n^t=\left[{p}_{n,1}^t, \ldots, {p}_{n,k}^t, \ldots, {p}_{n,K}^{t}\right]
\end{equation}
where ${p}_{n,k}^t$ denotes the power that the satellite $n$ allocates to the beam $k$ at time slot $t$.

\item Reward: To promote cooperation among the agents, they are awarded a global reward, which is defined as follows:
\begin{equation}
\begin{aligned}
&R^t=R^t\left(s_n^t, a_n^t\right)= \beta \frac{\sum_{c=1}^C Th_{n, c}^t}{Th_{\text{max}}}- \\
 &(1-\beta) \frac{ \left(\max \left\{\tau_{n,i}^{t}\right\}-\min \left\{\tau_{n,j}^{t}\right\}\right)}{J_{\text{max}}},  \forall i, j \in V_n
\end{aligned}
\end{equation}
\end{itemize}

\subsubsection{PA Based on MADDPG (MAPA)} \label{EE}
We employ the MADDPG algorithm, a multi-agent deep reinforcement learning method that merges the Actor-Critic (AC) and Deep Deterministic Policy Gradient (DDPG) algorithms, to solve the reformulated MDP problem. Since all onboard transmitters are deployed on the same satellite, each beam can observe the same state information. The decision for each beam is made by its own policy network, independent of other beams' actions. This process is known as decentralized decision-making. As shown in Fig.3, an actor network is deployed by each beam to choose actions, and a global critic network is used by the satellite to assess the action values.

\begin{figure}[htbp]
\centering
\includegraphics[height=6cm]{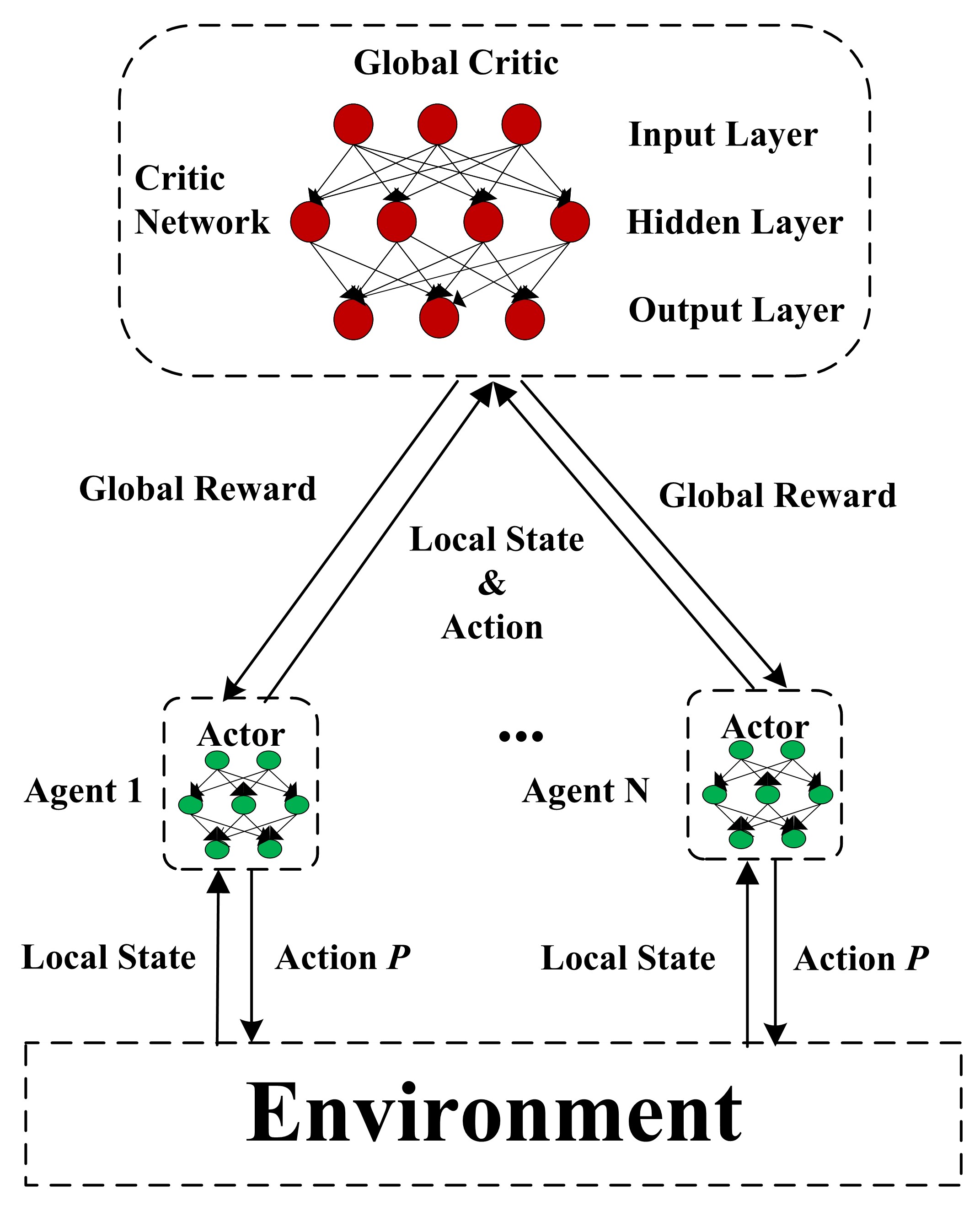}
\caption{The cooperative multi-agent actor-critic framework of \\the MAPA algorithm.}
\label{fig3}
\end{figure}
However, due to the cooperative and competitive relationships among the beams, an individual beam cannot reach an optimal solution without considering the actions of other beams. Therefore, after the beams make decentralized decisions, a policy evaluation network assesses the overall strategy of all beams on the satellite. The evaluation results are fed back to each beam's policy network to update the strategy parameters. Through continuous learning and iteration, the multiple beams can ultimately achieve a unique optimal solution. This process is known as centralized training.

In this work, each beam is treated as an agent and is equipped with its own actor network. The task of the actor network is to select actions, represented as:
\begin{equation}
\boldsymbol{a}_i^t=\pi_{\theta_i^t}(s_i^t)+ \mathcal{N}_0
\end{equation} 
where $\mathcal{N}_0$ represents exploration noise added to promote diverse action exploration. Concurrently, all agents execute their chosen actions $a_i$, and the environment responds with a new state $s'_i$ and an immediate reward $r_i$. The local critic network then calculates the $Q$ value, $y_i$, as
\begin{equation}
y_i^t = r_i^t + \gamma Q_{\phi'}(\{s_i^{t+1}\}, \{a_i^{t+1}\})|_{a_j^{t+1} = \pi_{\theta_{j'}}(s_j^{t+1})}
\end{equation}

Updates the global critic network parameters by minimizing the $Q$ value error:
\begin{equation}
L(\phi)=\frac{1}{|R|} \sum_j\left(y_j-Q_\phi\left(\left\{s_j\right\},\left\{a_j\right\}\right)\right)^2
\end{equation}

At the same time, each agent's local actor network parameters are updated using the policy gradient:
\begin{equation}
\nabla_{\theta^i} J \approx \frac{1}{|R|} \sum_j \nabla_{a_i} Q_{\phi}(\{s_j\}, \{a_j\})|_{a_i = \pi_{\theta^i}(s_i)} \nabla_{\theta^i} \pi_{\theta^i}(s_i)
\end{equation}

The specific procedure of the algorithm, called MAPA, is summarized in Algorithm 2.
\vspace{-0.1cm}
\begin{algorithm}
\caption{MADDPG for PA (MAPA)}
\begin{algorithmic}[1]

\State Initialize environment with satellites, beams, and cells, global power constraint $P$
\State Initialize actor network parameters $\theta_i$ for each agent $i$
\State Initialize global critic network parameters $\phi$
\State Initialize target network parameters $\theta_{i'} \leftarrow \theta_i$, $\phi' \leftarrow \phi$
\State Set learning rates $\alpha_{\theta}, \alpha_{\phi}$
\State Initialize replay buffer $\mathcal R$
\State Set the number of beams $K$, the number of time slots $T$
\State Set soft update parameter $\tau$

\For{each episode}
    \State Initialize state $s_i^0$ for each agent $i$
    \For{each time slot $t = 0, \ldots, T$}
        \For{each agent $i = 1, \ldots, K$}
            \State Select action $a_i^t$ according to (25)
        \EndFor
        \State Execute actions $\{a_i^t\}$ and observe new states $\{s_i^{t+1}\}$ and rewards $\{r_i^t\}$
        \State Store transition $(\{s_i^t\}, \{a_i^t\}, \{r_i^t\}, \{s_i^{t+1}\})$ in replay buffer $R$
        \State Sample a random minibatch of transitions $(\{s_j^t\}, \{a_j^t\}, \{r_j^t\}, \{s_j^{t+1}\})$ from $\mathcal R$

        \State \textbf{Global critic network update}:
        \For{each agent $i$}
            \State Set (26)
        \EndFor
        \State Update critic by minimizing the loss: (27)
        \State \textbf{actor network update}:
        \For{each agent $i$}
            \State Update actor using sampled policy gradient:
            \State (28)
        \EndFor

        \State \textbf{Soft update target networks}:
        \[
        \theta_{i'} \leftarrow \tau \theta_i + (1 - \tau) \theta_{i'};
        \phi' \leftarrow \tau \phi + (1 - \tau) \phi
        \]
    \EndFor
\EndFor

\end{algorithmic}
\end{algorithm}

\subsection{Complexity Analysis}
First, we analyze the complexity of using LSTM to predict the demand of a community. Assume that the LSTM network has $L$ layers, each layer has $H$ hidden units, and the time step is $T$. Then the complexity of LSTM is $\mathcal{O}(L_{lstm}\cdot H_{lstm}^2 \cdot T)$.

Secondly, we assess the complexity involved in solving sub-problem $P_1$. In the A3C algorithm adopted, each agent (satellite) has an Actor-Critic network, and there is a Critic network is centrally deployed. Assuming that the Actor network has $L_{\theta}$ layers, each layer has $H_{\theta}$ neurons, its complexity is $\mathcal{O}(L_{\theta} \cdot H_{\theta}^2)$. For the Critic network, assuming that there are $L_{\phi}$ layers, and each layer has $H_{\phi}$ neurons, the complexity of the local and global Critic network is: $\mathcal{O}(L_{\phi} \cdot H_{\phi}^2)$. Thus, the total complexity at each time step can be expressed as $\mathcal{O}(N\cdot(L_{\theta} \cdot H_{\theta}^2+2L_{\phi} \cdot H_{\phi}^2))=\mathcal{O}(N\cdot(L_{\theta} \cdot H_{\theta}^2+L_{\phi} \cdot H_{\phi}^2))$. In addition, the complexity of asynchronous global parameter update is $\mathcal{O}(I\cdot(L_{\theta} \cdot H_{\theta}^2+L_{\phi} \cdot H_{\phi}^2))$. As a result, the total complexity for solving sub-problem $P_1$ is $\mathcal{O}(I\cdot T \cdot N\cdot(L_{\theta} \cdot H_{\theta}^2+L_{\phi} \cdot H_{\phi}^2))$.  

Thirdly, the complexity of the MAPA algorithm for addressing sub-problem $P_2$ (i.e., the single-satellite PA problem) is analyzed. Similarly, assuming that the Actor network for each beam has $L_{{\theta}'}$ layers, each layer has $H_{{\theta}'}$ hidden neurons, and the input dimension is $D_A$, the Actor network’s complexity is expressed as $\mathcal{O}(\mathcal{B} \cdot L_{{\theta}'} \cdot H_{{\theta}'}^2)$. For the Critic network, assuming that the Critic network has $L_{{\phi}'}$ layers, and each layer has $H_{{\phi}'}$ neurons, the Critic network’s complexity is given by $\mathcal{O}(\mathcal{B} \cdot L_{{\phi}'} \cdot H_{{\phi}'}^2)$. Thus, the total complexity of sub-problem $P_2$ is $\mathcal{O}(T \cdot K \cdot \mathcal{B} \cdot(L_{{\theta}'} \cdot H_{{\theta}'}^2+ L_{{\phi}'} \cdot H_{{\phi}'}^2)$.

To summarize, the computational complexity of entire
algorithm BHPA-LBDP is $\mathcal{O}(L_{lstm}\cdot H_{lstm}^2 \cdot T)+\mathcal{O}(I\cdot T \cdot N\cdot(L_{\theta} \cdot H_{\theta}^2+L_{\phi} \cdot H_{\phi}^2)+\mathcal{O}(T \cdot K \cdot \mathcal{B} \cdot(L_{{\theta}'} \cdot H_{{\theta}'}^2+ L_{{\phi}'} \cdot H_{{\phi}'}^2)$.

\section{Performance Evaluation}
This section first describes the simulation environment and the parameters used for the simulations. Then, we present the simulation results to demonstrate the performance in DT-empowered multi-beam satellite network with the proposed algorithm BHPA-LBDP. 
\subsection{Simulation Parameter}
In this paper, we simulate a Ka-band forward BH and PA system. In both MA3C-BH and MAPA network architectures, the actor network comprises two hidden layers with 128 neurons each, while the critic network consists of two hidden layers with 256 neurons each. Adam is used as the solver for both networks. The constellation parameters and other main simulation parameters are summarized in Table II \cite{9968247}. 

We present the optimization effects of the proposed BH and PA Algorithm based on Load Balancing and Delay Fairness (named BHPA-LBDF, including MA3C-BH and MAPA) and compare it with four different BHPA schemes as follows:

1) Random Beam Hopping Fixed Power Allocation (RBH-FP):  In RBH-FP, each satellite randomly selects four cells to serve in each time slot, and distributes total power equally among the selected cells.

2) Random Beam Hopping Demand Based Power Allocation (RBH-DP): In RBH-DP, each satellite randomly selects four cells to serve in each time slot and allocates power according to the demand ratio of the selected cells.

3) Beam Hopping Fixed Power Allocation with Load Balancing and Delay Fairness (FPA): In FPA, based on load balancing across all satellites and service fairness within each satellite's coverage, four cells are selected for each satellite in each time slot, with total power divided equally among the selected cells.

4) Beam Hopping and Discretized Power Allocation Based on DT (DPA): In DPA, each satellite acts as an agent, completing BH and PA (discretized processing) to maximize system throughput and delay fairness.

\begin{table}[htbp]
\centering
\caption{System Parameters}
\begin{tabularx}{\columnwidth}{@{}X|l@{}}
\toprule
\textbf{Parameters} & \textbf{Values} \\
\hline \textbf{SYSTEM PARAMETERS} &  \\
Satellite altitude & 780 km \\
Ka-band frequency & 20 GHz \\
Number of satellites, $N$ & 12 \\
The number of cells served by per satellite, $C$ & 19 \\
Total number of cells covered by all satellites, $V$ & 168 \\
The coverage radius of a cell, $R$ & 39 km \\
Number of beams per satellite, $K$ & 4 \\
The available bandwidth for the system, $B$ & 100 MHz \\
Satellite's total power, $P$ & 39 dBW \\
The maximum power for each beam, $P_{max}$ & 30 dBW \\
TAperture radius of the antenna  & 0.15 m \\
The satellite beams' 3 dB beamwidth & 3$^{\circ}$ \\
Maximum gain of the transmit antenna & 35.9 dBi \\
Receiving antenna gain at the terminal, $G_r$ & 0 dBi \\
Temperature of noise, $T_{rx}$ & 300 K \\
Time slot duration, $T_{\text{slot}}$ & 2 ms \\
The period of BHTP, $T_H$ & 64 $\times$ $T_{\text{slot}}$ \\
Weighting factors, $\alpha, \beta$ & $\alpha=\beta=0:0.1:1$ \\
\hline \textbf{MA3C-BH TRAINING PARAMETERS} &  \\
Training Episode & 6000 \\
Actor network learning rate, $\alpha_{\theta}$ & $0.00001$ \\
Critic network learning rate, $\alpha_{\phi}$ & $0.0001$ \\
Discount factor, $\gamma$ & 0.99 \\
Number of Threads, $I$ & 16 \\
Optimizer & Adam \\
\hline \textbf{MA-PA TRAINING PARAMETERS} &  \\
Training Episode & 6000 \\
Soft Update Parameter, $\tau$ & 0.001 \\
Replay Buffer size, $\mathcal R$ & 1000000 \\
Minibatch size, $|\mathcal{B}|$ & 64 \\
Discount factor, $\gamma$ & 0.99 \\
Actor network learning rate $\alpha_{\theta}$ & $0.00001$ \\
Critic network learning rate $\alpha_{\phi}$ & $0.0001$ \\
Exploration Noise $\mathcal{N}_0$ & 0.2 \\
Optimizer & Adam \\
\bottomrule %[2pt]     
\end{tabularx}
\end{table}

\vspace{-0.2cm}
\subsection{Convergence Analysis}
To illustrate the convergence of the BHPA-LBDF method, we present the episode reward and episode loss from one of the training scenarios with $\alpha = \beta = 0.5$, as depicted in Fig. 4. As shown in Fig. 4(a), the reward increases with the training episodes and converges after 1200 episodes. Fig. 4(b) shows that the episode loss remains stable and converges to $10^{-4}$ after approximately 4000 episodes. This demonstrates that the convergence of BHPA-LBDF is guaranteed under the parameter settings listed in Table II. Next, we will compare the performance of BHPA-LBDF with the four aforementioned schemes. The performance metrics include throughput, load balance, queueing delay and delay fairness.
\vspace{-0.4cm}
\begin{figure}[htbp]
\centering
\subfloat[]{
		\includegraphics[height=3.4cm]{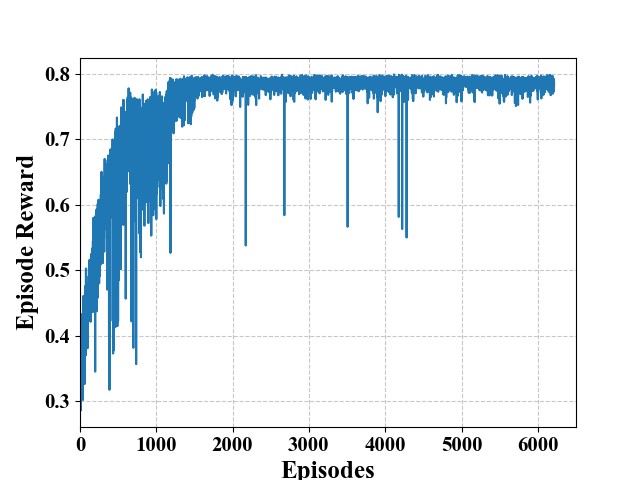}}
\subfloat[]{
		\includegraphics[height=3.4cm]{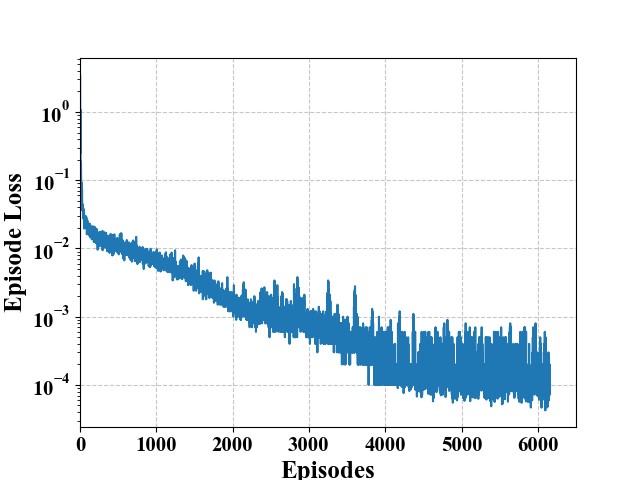}}
\caption{An illustration of (a) episode reward and (b) loss during training \\($\alpha$ = $\beta$ = 0.5)}
\label{fig4}
\end{figure}
\vspace{-0.4cm}
\subsection{Performance Analysis}
\vspace{-0.4cm}
\begin{figure}[htbp]
\centering
\subfloat[]{
		\includegraphics[height=3cm]{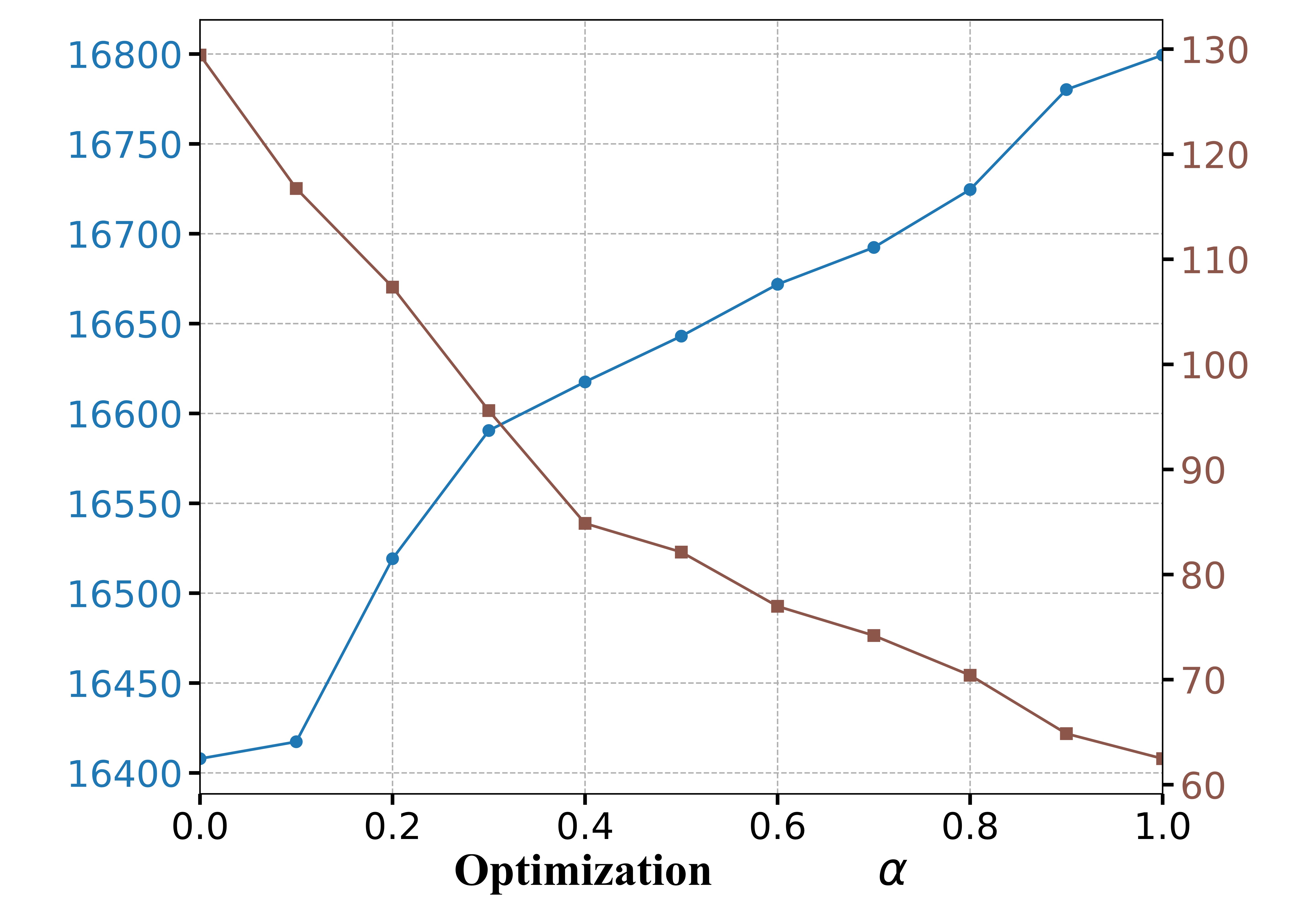}}
\subfloat[]{
		\includegraphics[height=3.3cm]{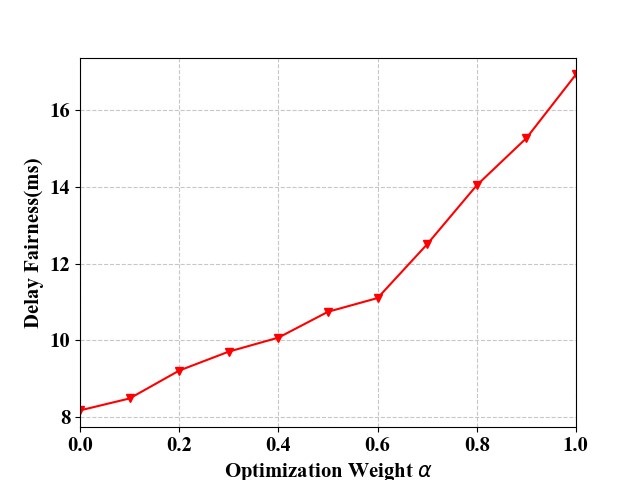}}
\caption{The performance of (a) traffic load-traffic load gap and (b) delay fairness under different optimization weight $\alpha$.}
\label{fig5}
\end{figure}

\subsubsection{Performance Under Different Optimization Weights} The optimization weight $\alpha$ reflects the trade-off between load balance $Q$ and delay fairness $J$. As shown in Fig. 5, the performance of traffic load, traffic load gap and delay fairness changes with $\alpha$. Both traffic load and delay fairness increase as $\alpha$ rises from $0$ to $1$ since a larger $\alpha$ focuses more on load balancing. This reduces the load gap, thereby improving delay fairness. A higher delay fairness indicates greater inequity in resource allocation. To strike a balance between load and delay fairness, we use the trained model with $\alpha = 0.5$ for comparison with other approaches.
\vspace{-0.3cm}
\begin{figure}[htbp]
\centering
\subfloat[]{
		\includegraphics[height=3.4cm]{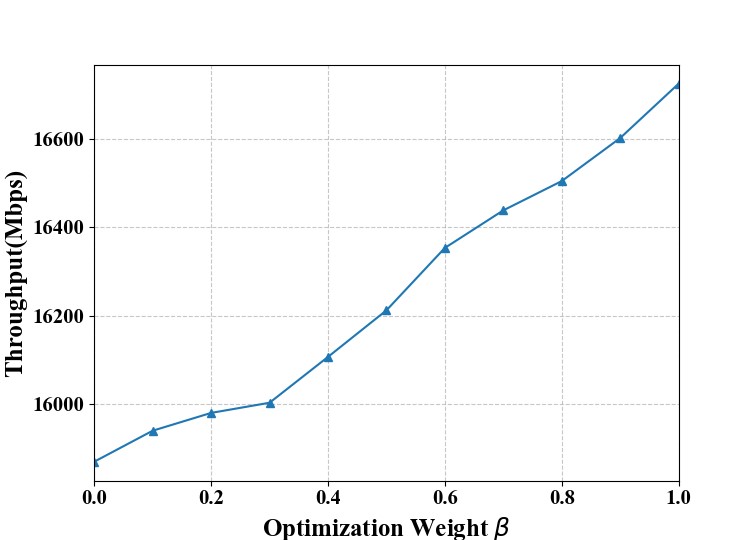}}
\subfloat[]{
		\includegraphics[height=3.4cm]{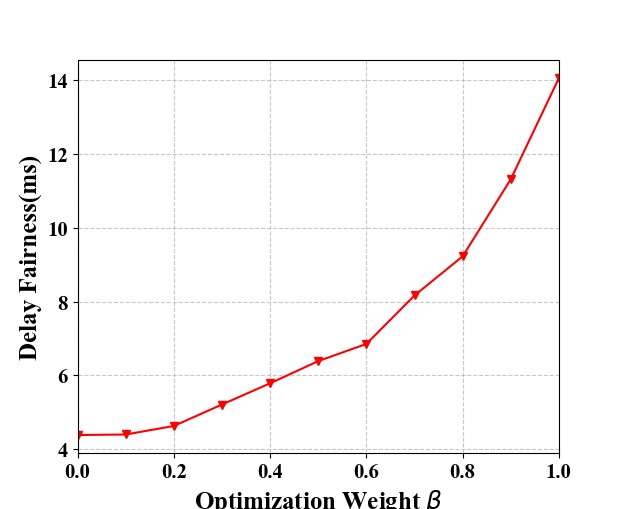}}
\caption{The performance of (a) throughput and (b) delay fairness \\under different optimization weight $\beta$.}
\label{fig6}
\end{figure}

Similarly, the optimization weight $\beta$ reflects the trade-off between throughput and delay fairness. As shown in Fig. 6, the performance of both throughput and delay fairness improves as $\beta$ increases from $0$ to $1$. To achieve a better balance between throughput and delay fairness, we use the trained model with $\beta = 0.5$ for comparison with other approaches.

\subsubsection{Performance of Load Balancing}
Fig. 7 shows the traffic load of each satellite after BH using different algorithms. The red box represents the BH strategy using MA3C algorithm, which considers load balancing and delay fairness, achieving the smallest load difference of 82.13 Mbps between satellites. Notably, due to uneven traffic distribution, MA3C algorithm improves load balancing by 75.95\%, 43.45\%, 60.95\%, and 73.11\% compared to the Greedy BH (G-BH), Random BH (R-BH), Periodic BH (P-BH), and Queue-Length Priority BH (Q-BH) methods, respectively.
\begin{figure}[htbp]
\centering
\includegraphics[width=8.5cm]{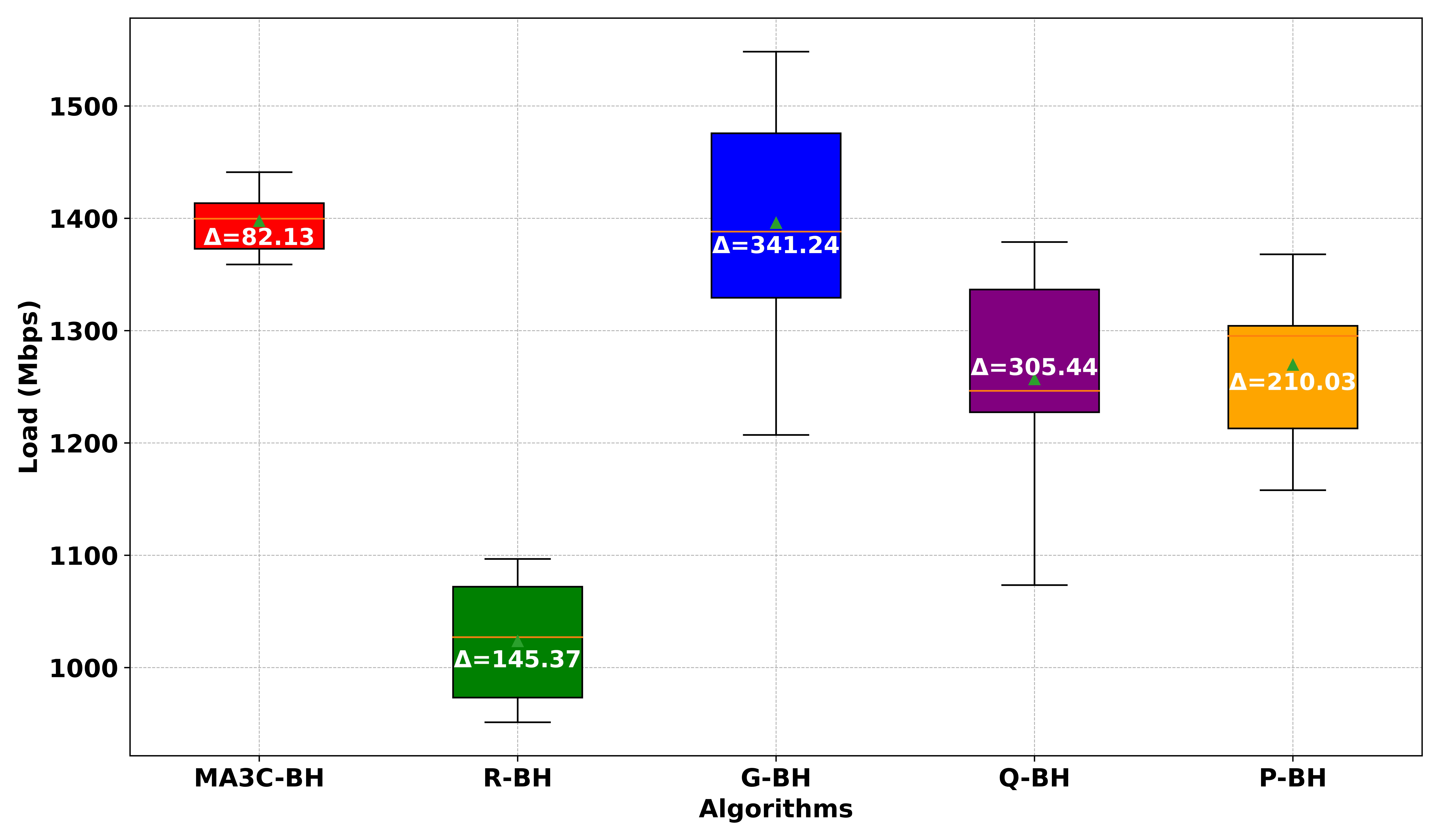}
\caption{ Results of load balancing after beam hopping for different methods \\(the proposed MA3C-BH with $\alpha$= 0.5).}
\label{fig7}
\end{figure}

\subsubsection{Performance of Throughput}
Fig. 8 shows the throughput versus total traffic demand using different methods, and Fig. 9 shows throughput of 12 satellites among different algorithms. The total traffic demand represents the cumulative demand of all cells. The proposed method, BHPA-LBDP, consistently outperforms other algorithms. This improvement occurs because agents accumulate experience during training, eventually learning to fully utilize the beam's degrees of freedom in time, space, and power. Specifically, compared to RBH-FP, RBH-DP, FPA, and DPA, the throughput of the proposed method improves by 96.7\%, 46.2\%, 10.2\%, and 5.6\%, respectively. 

As the total traffic demand increases, the throughput of the BHPA-LBDP, FPA, and DPA algorithms consistently surpasses that of the RBH-FP and RBH-DP algorithms. This is mainly because BHPA-LBDP, FPA, and DPA use reinforcement learning to select service cells for BH based on cell demand, whereas RBH-FP and RBH-DP randomly select cells for service. The throughput of RBH-DP is always higher than that of RBH-FP, especially as the total demand increases, with the gap between them gradually widening. This indicates that intelligent PA significantly enhances overall throughput. This is also validated by the comparison of throughput between the BHPA-LBDP and FPA algorithms.

When the total demand is between 8000 and 20000 Mbps, the throughput of the FPA and DPA algorithms shows little difference, with FPA being slightly higher than FPA. This is because both algorithms perform intelligent BH based on cell demand, and the capacity of each beam exceeds the traffic demand of each cell, resulting in a relatively minor impact from PA on overall throughput. However, when the demand exceeds 20000 Mbps, the throughput of DPA becomes significantly higher than that of FPA. This is mainly because, with increasing total demand, DPA can flexibly allocate limited power resources to different cells. Nevertheless, due to the discretization of power by DPA and its joint decision on BHPA within DT, its throughput is always lower than that of our proposed BHPA-LBDP algorithm.
\vspace{-0.3cm}
\begin{figure}[htbp]
\centering
\includegraphics[width=8.5cm]{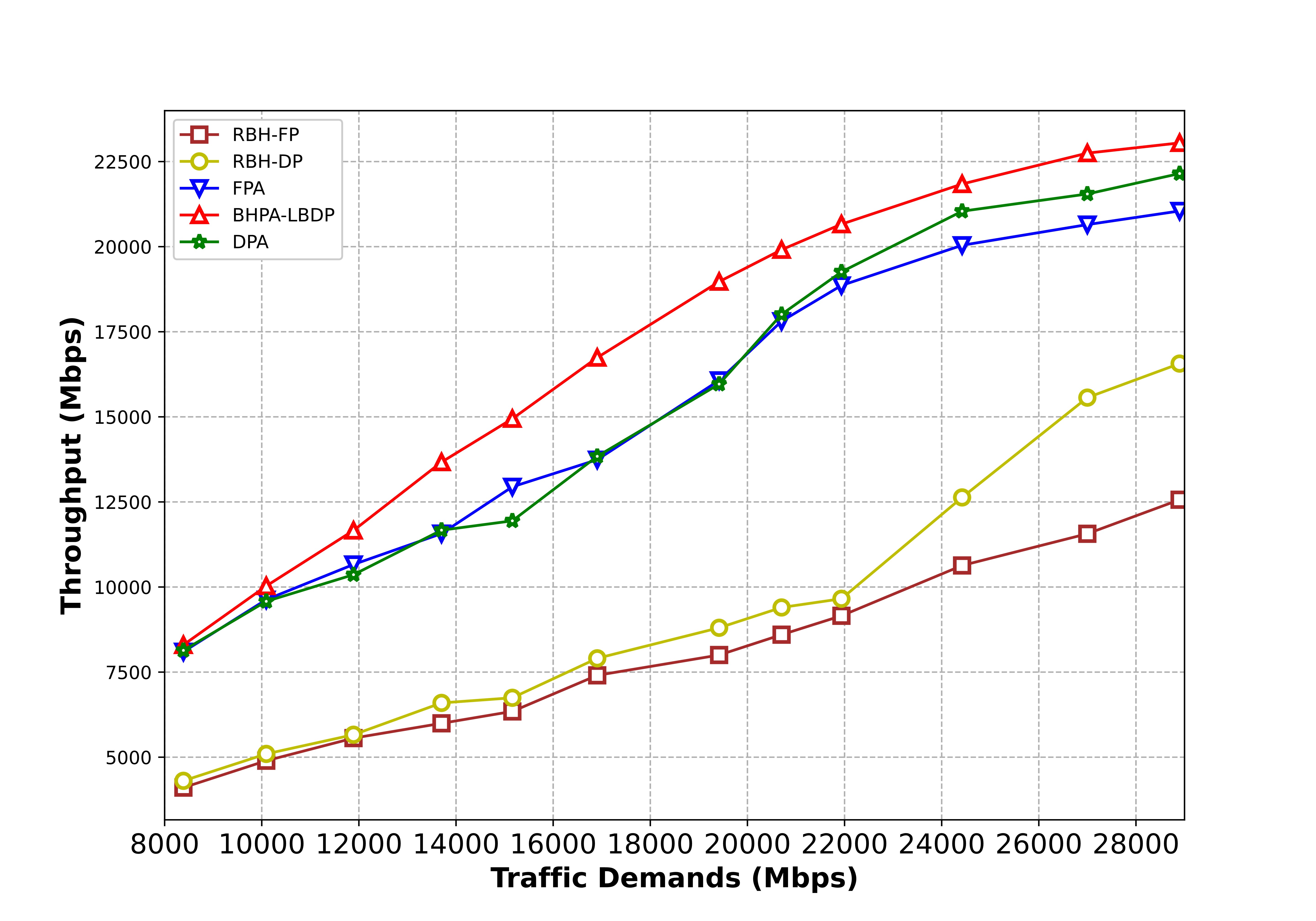}
\caption{The comparison of throughput versus total traffic demand for different methods (the proposed BHPA-LBDP with $\alpha$=$\beta$ = 0.5).}
\vspace{-0.8cm}
\label{fig8}
\end{figure}

\begin{figure}[htbp]
\centering
\includegraphics[width=8.5cm]{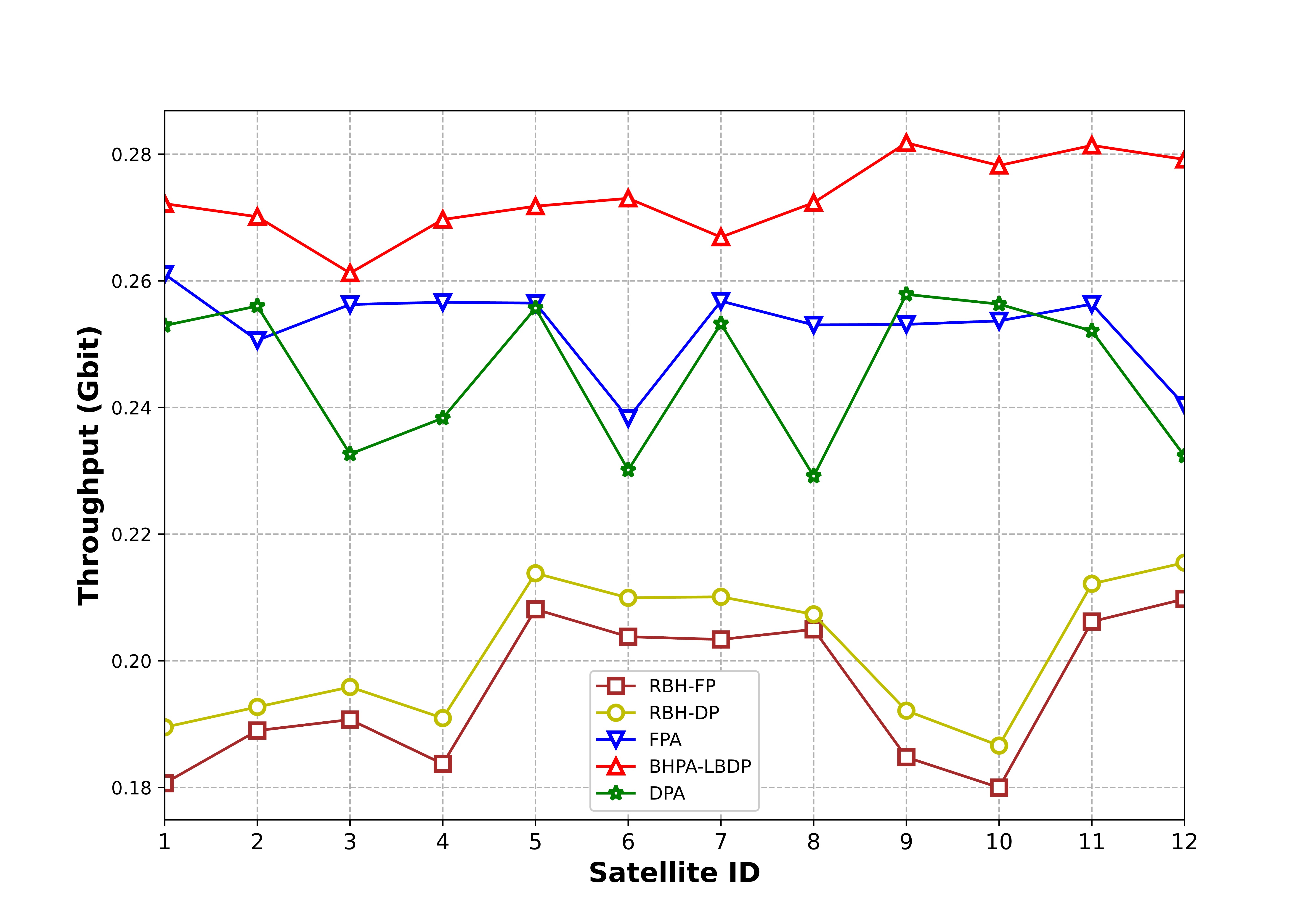}
\caption{The throughput of 12 satellites among different algorithms. \\(the proposed BHPA-LBDP with $\alpha$=$\beta$ = 0.5).}
\label{fig9}
\end{figure}
\subsubsection{Performance of Delay Fairness}
Fig. 10 illustrates the relationship between the delay fairness of various methods and the total traffic demand. Fig. 11 shows the average queuing delay among different satellites for the five algorithms.

With the increase in total traffic demand, delay fairness improves for all methods, as the limited beam resources are insufficient to handle the excessive traffic, resulting queue congestion. Throughout this process, the queueing delay of intelligent BH algorithms (BHPA-LBDP, FPA, and DPA) remains consistently lower than that of the random BH algorithm. Our proposed BHPA-LBDP algorithm also maintains a lowest queue delay. Compared to the RBH-FP, RBH-DP, FPA, and DPA algorithms, the average queue delay is reduced by 46\%, 43.23\%, 10.8\%, and 20.2\%, respectively. Additionally, comparing the three algorithms, i.e., BHPA-LBDP, FPA, and DPA, reveals that PA decision-making based on actual channel information on the satellite has significant advantages in improving overall performance.

\begin{figure}[htbp]
\centering
\includegraphics[width=8cm]{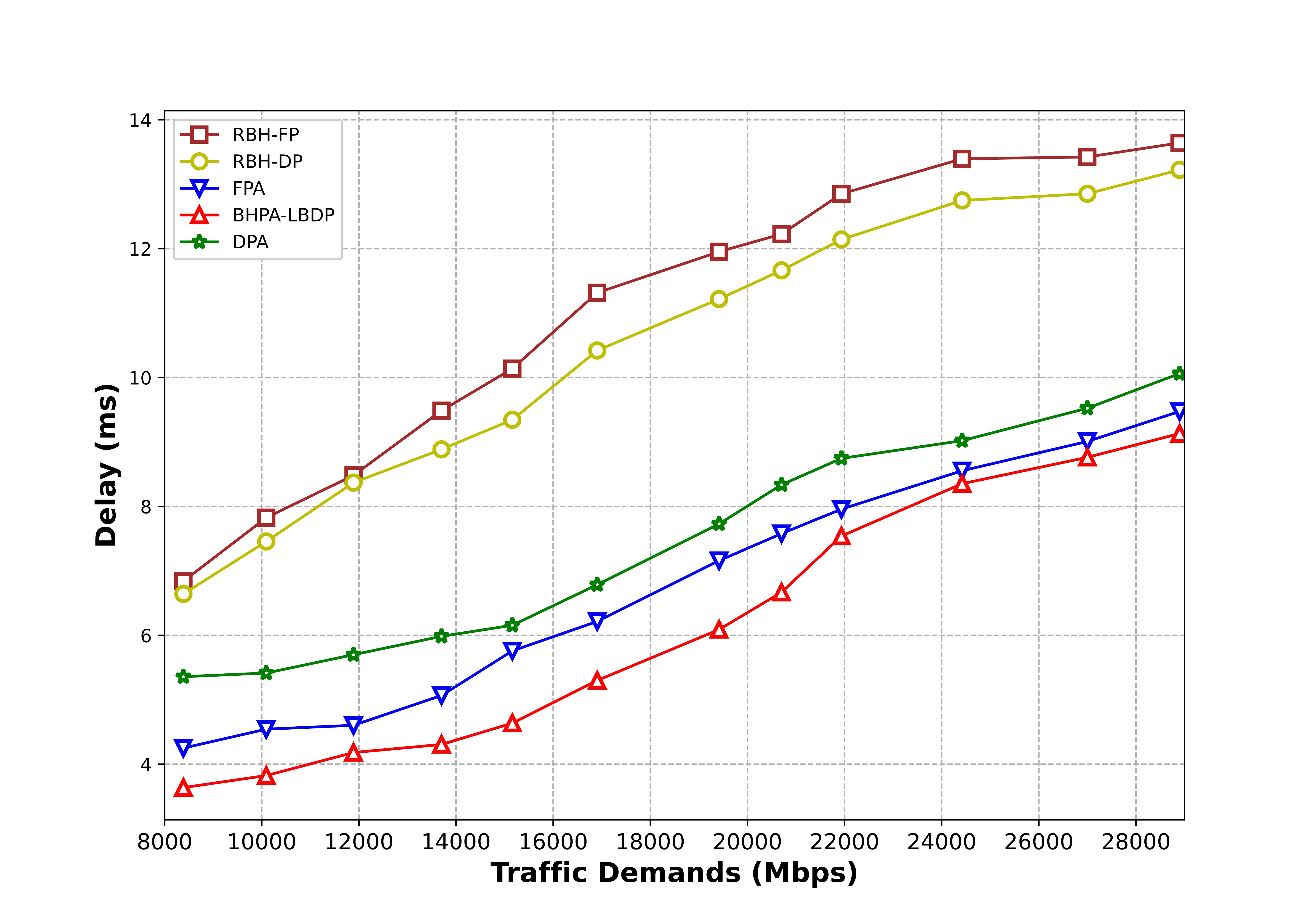}
\caption{ The delay fairness versus total traffic demand of different methods.\\ (the proposed BHPA-LBDP with $\alpha$=$\beta$ = 0.5).}
\vspace{-0.5cm}
\label{fig10}
\end{figure}

\begin{figure}[htbp]
\centering
\includegraphics[width=8cm]{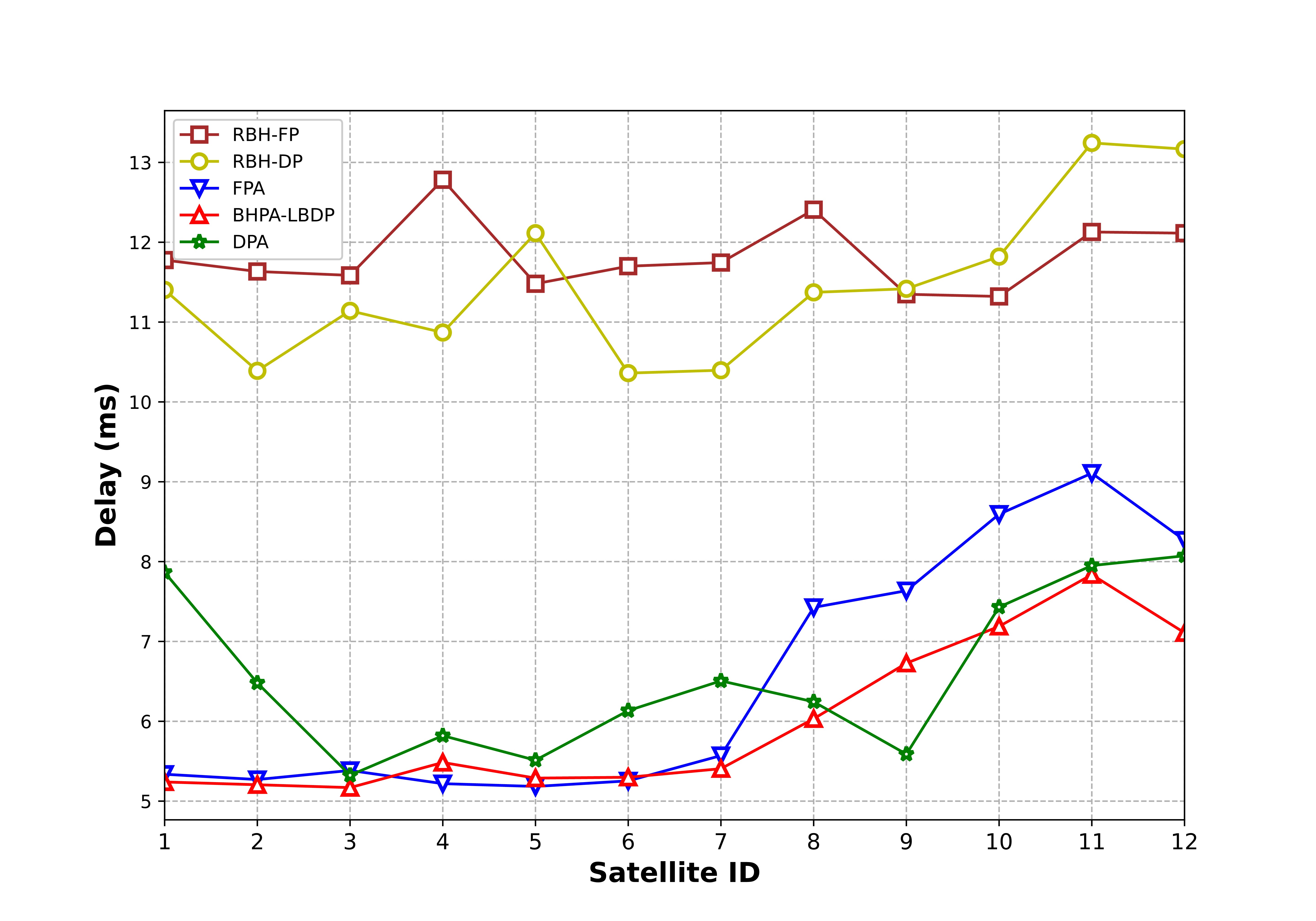}
\caption{ The delay fairness of 12 satellites among different algorithms. \\(the proposed BHPA-LBDP with $\alpha$=$\beta$ = 0.5).}
\label{fig11}
\end{figure}

\section{Conclusion}
This paper investigates the coordinated BH and PA problem in multi-beam LEO satellite constellations to align limited satellite resources with uneven service demands. Initially, a digital-twin-based framework for coordinated BH and resource optimization among multiple satellites is established. Acknowledging the NP-hard nature of the optimization task, the problem is decomposed into two subsidiary problems. First, leveraging the multiple coverage offered by LEO satellite constellations, cloud center is employed to select spatially isolated BH patterns for each satellite. This approach balances inter-satellite load demands, ensures equitable service provisioning, and mitigates interference. Second, PA is dynamically managed through a collaborative competitive mechanism among beams within each satellite, catering to real-time needs. Finally, simulations validate that this methodology outperforms existing BHPA schemes in terms of thoughput and delay fairness.

In the future, we aim to develop a comprehensive resource allocation strategy that integrates the optimization of BH, spectrum, time, and power resources to maximize long-term user data rates and minimize system power consumption. We will focus on algorithms that leverage the combination of digital twins and artificial intelligence to enable dynamic and intelligent management of onboard resources, ensuring optimal system performance in dynamic environments. Additionally, we plan to introduce intelligent prediction mechanisms to proactively identify and address potential interference and resource bottlenecks, thereby enhancing overall system efficiency and reliability. We anticipate that these improvements will significantly improve the effectiveness of our proposed joint BH and multi-dimensional resource allocation framework, creating a more efficient and energy-saving communication environment for LEO satellite networks.

\bibliography{ref.bib}
\bibliographystyle{ieeetr}

\end{document}